\documentclass[twocolumn]{aastex631}
\usepackage{xspace}
\usepackage{amsmath}
\usepackage{booktabs} % For better-looking tables
\newcommand{\prompt}

\newcommand{\Msun}{M$_{\odot}$\xspace}
\newcommand{\Msunyr}{M$_{\odot}$ yr$^{-1}$\xspace}

\newcommand{\epsb}{$\epsilon_B$\xspace}
\newcommand{\epse}{$\epsilon_e$\xspace}
\newcommand{\lumunit}{erg s$^{-1}$ Hz$^{-1}$\xspace}
\newcommand{\masslossunit}{M$_{\odot}$ yr$^{-1}$ (100 km s$^{-1}$)$^{-1}$\xspace}

\hypersetup{linkcolor=blue,citecolor=blue,filecolor=blue,urlcolor=blue}
\shorttitle{Radio Ia-CSM}
\shortauthors{Griffith et al.}
\begin{document}

\title{A Late-time Radio Survey of Type Ia-CSM Supernovae with the Very Large Array}

\author[0000-0000-0000-0000]{Olivia Griffith}
\affiliation{Center for Data Intensive and Time Domain Astronomy, Department of Physics and Astronomy, Michigan State University, East Lansing, MI 48824, USA}

\author[0009-0000-2871-9330]{Grace Showerman}
\affiliation{Center for Data Intensive and Time Domain Astronomy, Department of Physics and Astronomy, Michigan State University, East Lansing, MI 48824, USA}

\author[0000-0002-4781-7291]{Sumit K. Sarbadhicary}
\affiliation{Department of Physics, The Ohio State University, Columbus, Ohio 43210, USA}
\affiliation{Center for Cosmology \& Astro-Particle Physics, The Ohio State University, Columbus, Ohio 43210, USA}
\affiliation{Department of Astronomy, The Ohio State University, Columbus, Ohio 43210, USA}
\affiliation{Department of Physics and Astronomy, The Johns Hopkins University, Baltimore, MD 21218 USA}

\author[0000-0002-8400-3705]{Chelsea E. Harris}
\affiliation{Center for Data Intensive and Time Domain Astronomy, Department of Physics and Astronomy, Michigan State University, East Lansing, MI 48824, USA}

\author[0000-0002-8400-3705]{Laura Chomiuk}
\affiliation{Center for Data Intensive and Time Domain Astronomy, Department of Physics and Astronomy, Michigan State University, East Lansing, MI 48824, USA}

%\affiliation{Astronomical Science Program, Graduate Institute for Advanced Studies, SOKENDAI, 2-21-1 Osawa, Mitaka, Tokyo 181-8588, Japan}
%\affiliation{School of Physics and Astronomy, Monash University, Clayton, VIC 3800, Australia}

\author[0000-0003-1546-6615]{Jesper Sollerman}
\affiliation{The Oskar Klein Centre, Department of Astronomy, Stockholm University, AlbaNova, SE-10691, Stockholm, Sweden}

\author[0000-0002-3664-8082]{Peter Lundqvist}
\affiliation{The Oskar Klein Centre, Department of Astronomy, Stockholm University, AlbaNova, SE-10691, Stockholm, Sweden}

\author[0000-0002-8079-7608]{Javier Mold\'{o}n}
\affiliation{CSIC, Instituto de Astrof\'isica de Andaluc\'ia,  Glorieta de la Astronom\'ia S/N, E-18008, Granada, Spain}

\author[0000-0001-5654-0266]{Miguel P\'erez-Torres} 
\affiliation{CSIC, Instituto de Astrof\'isica de Andaluc\'ia,  Glorieta de la Astronom\'ia S/N, E-18008, Granada, Spain}
\affiliation{School of Sciences, European University Cyprus, Diogenes street, Engomi, 1516 Nicosia, Cyprus}

\author[0000-0002-7252-3877]{Erik C. Kool}
\affiliation{The Oskar Klein Centre, Department of Astronomy, Stockholm University, AlbaNova, SE-10691, Stockholm, Sweden}
\affiliation{Finnish Centre for Astronomy with ESO (FINCA), University of Turku, 20014 Turku, Finland}
\affiliation{Department of Physics and Astronomy, University of Turku, 20014 Turku, Finland}

\author[0000-0003-1169-1954]{Takashi J. Moriya}
\affiliation{National Astronomical Observatory of Japan, National Institutes of Natural Sciences, 2-21-1 Osawa, Mitaka, Tokyo 181-8588, Japan}
\affiliation{Graduate Institute for Advanced Studies, SOKENDAI, 2-21-1 Osawa, Mitaka, Tokyo 181-8588, Japan}
\affiliation{School of Physics and Astronomy, Monash University, Clayton, VIC 3800, Australia}
%\suppressAffiliations

\correspondingauthor{Olivia Griffith}
\email{griff717@msu.edu}

\begin{abstract}
Type Ia-CSM supernovae (SNe) are a rare and peculiar subclass of thermonuclear SNe characterized by emission lines of hydrogen or helium, 
indicative of high-density circumstellar medium (CSM). Their implied mass-loss rates of $\sim 10^{-4}-10^{-1}$ \Msunyr (assuming $\mathrm{ \sim 100 \ km\ s^{-1}}$ winds) from optical observations are generally in excess of values observed in realistic SN Ia progenitors. In this paper, we present an independent study of CSM densities around a sample of 29 archival Ia-CSM SNe using radio observations with the Karl G.\ Jansky Very Large Array at 6 GHz. Motivated by the late ($\sim$2 yr) radio detection of the Ia-CSM SN 2020eyj, we observed old ($>$1 yr) SNe, when we are more likely to see the emergent synchrotron emission that may have been suppressed earlier by free-free absorption by the CSM. We do not detect radio emission down to 3$\sigma$ limits of $\sim$35 $\mu$Jy in our sample. The only radio-detected candidate in our sample, SN 2022esa, was likely mis-classified as a Ia-CSM with early spectra, and appears more consistent with a peculiar Ic based on later-epochs. Assuming wind-like CSM with temperatures between $2 \times 10^4$ K and $10^5$ K, and magnetic field-to-shock energy fraction \epsb = $0.01-0.1$,  the radio upper limits rule out  mass-loss rates between $\sim 10^{-4}-10^{-2}$ \masslossunit. This is somewhat in tension with the estimates from optical observations, and may indicate that more complex CSM geometries and/or lower values of $\epsilon_B$ may be present. %%Our published catalog will be a valuable resource for investigating SN Ia-CSM models, esspecially those with complex geometries and particle acceleration models that may alleviate the radio-optical tension.
%and the detection of radio emission from SN~2022esa

\end{abstract}

\keywords{}


\section{Introduction}
\label{sec:1}
The progenitor scenario(s) for Type Ia supernovae (SNe Ia) remains one of the biggest open questions in astrophysics. The two broad scenarios explored by the community have been the single degenerate scenario, involving a white dwarf accreting from a non-degenerate main-sequence star, red giant, or helium star companion \citep[e.g.][]{whelan73, nomoto82}, and the double-degenerate scenario, involving two white dwarfs in a binary system \citep[e.g.][]{tutukov79, iben84, webbink84, shen18}.  Unfortunately, observations have yet to establish which combination of these scenarios are contributing to the observed SN Ia population in the universe \citep[see e.g][for recent reviews]{Jha2019, Liu2023, Ruiter2025}. Considering the importance of SNe Ia as the the primary producers of Fe-group elements \citep{raiteri96}, as drivers of chemical enrichment in the universe \citep{matteucci01}, and as standard candles \citep[e.g.,][]{pskovskii77,phillips93} for precise astrophysical distance measurements \citep{phillips93, riess98, perlmutter99, phillips99, burns18}, the resolution of the SN Ia progenitor problem remains a high-priority goal for our field.

Here we focus on a particularly intriguing and rare sub-class of SNe Ia that have eluded explanation -- SNe Ia-CSM \citep{ Silverman2013}. They are characterized by strong narrow emission lines of hydrogen which have been attributed to circumstellar material (CSM) produced by mass-loss from the progenitor. SNe Ia-CSM also appear to be more luminous than normal SNe Ia, and have a preference for star-forming host galaxies \citep{Sharma2023}. They have been considered strong candidates for the single-degenerate channel, since white dwarfs with red-giant or asymptotic giant branch (AGB) star companions can have strong winds \citep[e.g.][]{Han2006, Dilday12}. Some well-known examples of Ia-CSM SNe include PTF11kx \citep{Dilday12}, SN 2002ic \citep{Hamuy2003} and SN 2005gj \citep{Aldering2006}. Many of them were previously confused with Type IIn SNe owing to dilution and blending of key spectral features by the bright continuum \citep{Fox2015, Leloudas2015, Inserra2016}. In very rare cases, CSM interaction can appear at late-times in SNe Ia due to distant, detached shells of CSM \citep[e.g.,][]{Graham2019, Wang2024, Terwel2025, Mo2025}.

A major conundrum for SNe Ia-CSM is that their implied mass-loss rates from the H emission lines are typically in the range of $10^{-4}-10^{-2}$ \Msun yr$^{-1}$ \citep[assuming wind velocities of $\mathrm{100 \ km\ s^{-1}}$, e.g.][]{Aldering2006, Dilday12, Silverman2013, Sharma2023}. These are $2-4$ orders of magnitude higher than observed in realistic single-degenerate progenitors with the highest mass-loss rates, such as symbiotic systems with AGB companions \citep[e.g.][]{Seaquist1990, Seaquist1993, Chomiuk2012_v407cyg}. Other scenarios such as dense shells of swept-up mass from recurrent novae \citep[e.g.][]{Moore2012, Darnley2019}, or common-envelope shells preceding formation of double-degenerate systems \citep[e.g.][]{Livio2003} have also been invoked, but a clear connection between observations and progenitors is lacking so far for SNe Ia-CSM. 

An orthogonal way of studying CSM properties is with radio observations, which trace synchrotron emission from SN shocks interacting with the ambient medium \citep{Chev82}. Given the observed luminosity or upper limits, and some assumptions about the fractional shock energy shared with the electrons and magnetic fields, reasonable estimates of the CSM density can be obtained. While radio emission has been observed in many core-collapse SNe interacting with the dense CSM around their massive progenitors \citep[e.g.,][]{Weiler2002, Chevalier2006, Soderberg2012}, similar emission has been elusive in Type Ia SNe. Their non-detections have provided strong evidence that most SNe Ia have low-density environments, ruling out symbiotic companions, and even a substantial parameter space of nova-driven and accretion-disk-driven winds \citep[e.g.,][]{panagia2006,chomiuk2012, chomiuk16, Horesh2012, Lundqvist2020, Harris23}.

In contrast to the general SN Ia population, there are fewer constraining radio observations of Ia-CSM SNe. \cite{chomiuk16} carried out the largest radio survey of SNe Ia, in which there were six Ia-CSM SNe that all showed non-detections. However, these observations were generally taken within a few months of the explosion, and therefore could have missed the radio emission due to synchrotron self-absorption and free-free absorption by the ionized CSM. This is clearly seen in radio observations of their closest counterparts, Type IIn SNe, whose radio light curves peak a few years after explosion \citep{Chandra2018, Bietenholz2021, Sfaradi2025}. Unfortunately, very few late-time radio observations of SNe Ia-CSM exist on these timescales to detect emission and independently constrain the properties of the high density CSM. 

In this paper, we present the first and largest ``late-time" (i.e. $>$1 yr after explosion) radio survey of SNe Ia-CSM with the Karl G.\ Jansky Very Large Array (VLA). The survey was inspired by the discovery of radio emission from the He-rich Ia-CSM, SN 2020eyj, which at a distance of $\sim$131 Mpc produced bright radio emission ($\sim$10$^{27}$ erg s$^{-1}$ Hz$^{-1}$) at 5--6 GHz on timescales of $\sim$1.5--2 yrs after explosion \citep{Kool23+}. This is the first and only thermonuclear SN to date with detected radio emission, and together with the late-time radio emission in Type IIn SNe \textbf{\citep{Bietenholz2021}}, encourages a similar search of Ia-CSM SNe, sufficiently late in time to detect any emergent emission. Here, we target 29 archival SNe Ia-CSM with ages of at least 1 yr, and analyze them with synchrotron emission models of interaction with a $\rho \propto r^{-2}$ wind-like CSM in order to obtain constraints on their mass-loss rates, enabling comparison with values deduced from optical observations. 
%The exact ages and distances for each SN are recorded in Table \ref{tab:observations}.

The paper is organized as follows: Section \ref{sec:obs} describes the VLA observations of our 29 targets. Section \ref{sec:3} discusses the radio light curve model used for deriving mass-loss rates (or upper limits) from the observed luminosities. Section \ref{sec:results} goes over the results of our observations and measurements of mass-loss rates, and Section \ref{sec:discussion} discusses these in the context of mass-loss rate measurements from other wavelengths, and  briefly considers the detectability of shell-like CSM. We conclude in Section \ref{sec:6}.

\section{Observations} \label{sec:obs}
In this section, we describe the radio observations conducted on our SN Ia-CSM sample. We observed 29 SNe Ia-CSM at 6 GHz with the VLA in the A configuration during July--Aug 2023 as part of the program VLA/23A-328 (PI: Sarbadhicary).
%\footnote{Four of the SNe proposed in our program -- SN 2005jg, 2005JG, 2013I, 2021admm -- were part of a scheduling block that did not get observed during the 2023A semester as a result of C-priority.} 
The sample was selected from published SNe Ia-CSM in \cite{Silverman2013} and \cite{Fox2015}, and transients classified as ``Ia-CSM" in the Transient Name Server for SNe Ia discovered after 2016. We excluded SNe Ia-CSM that are outside the declination limit of the VLA,  selecting targets at $\delta > -35^{\circ}$. Our final sample consists of SNe with ages of 1 to 25 years, which are located at distances ranging from 90 to 900 Mpc (Table \ref{tab:observations}). These distances were calculated with the redshift values given in the Transient Name Server along with the assumption of $H_0$=70 km s$^{-1}$  Mpc$^{-1}$.
%%$\Omega_M$=0.3 and %%$\Omega_{\Lambda}$=0.7}.
%While our sample is not statistically complete or volume-limited, it is large enough to test what fraction of SNe Ia-CSM have late-time radio emission as a result of high-density CSM implied by optical observations. 
 \begin{table*}
  \centering
  \caption{Radio Observations of our SN Ia-CSM sample.}
  \begin{tabular}{llrrrrrrr}
  \toprule
  SN Name & Discovery & Observation & Time Since & RA & Dec & Distance & 3$\sigma$ Flux & 3$\sigma$ Luminosity \\
  & Date & Date & Explosion & (J2000) & (J2000) & & Upper Limit & Upper Limit \\
  & (UT) & (UT) & (Years) & (h:m:s) & ($^{\circ}:^{\prime}:^{\prime\prime}$) & (Mpc) & ($\mu$Jy) & ($10^{27}$ erg s$^{-1}$ Hz$^{-1}$) \\
  \midrule
  SN 1999E & 99/01/15 & 23/08/20 & 24.6 & 13:17:16.37 & $-$18:33:13.4& 109.1 &$<$36 & $<$0.52\\
  SN 2002ic & 02/11/13 & 23/08/28 & 20.8 & 01:30:02.55 & 21:53:06.9& 296.8 & $<$29 & $<$3.1\\
  SN 2008cg & 08/05/05 & 23/08/25 & 15.3 & 15:54:15.15 & 10:58:25.0 & 159.3 & $<$41 & $<$1.3 \\
  SN 2009in & 09/08/25 & 23/08/27 & 14.0 & 23:22:35.32 & $-$13:05:43.6& 103.3 & $<$33 & $<$0.43 \\
  PTF10htz & 10/04/03 & 23/08/16 & 13.4 & 13:08:37.52 & 79:47:13.2& 153.9 & $<$48 & $<$1.4\\
  SN 2011dz & 11/06/26 & 23/08/25 & 12.2 & 16:12:44.82 & 28:17:03.2 & 106.9 & $<$37 & $<$0.51 \\
  SN 2011jb & 11/11/28 & 23/08/13 & 11.7 & 11:37:04.80 & 15:28:14.2& 382.5 & $<$34 & $<$5.9 \\
  CSS120327 & 12/03/27 & 23/08/20 & 11.4 & 11:05:20.08 & $-$01:52:05.2& 415.4 & $<$39 & $<$8.0 \\
  SN 2013dn & 13/06/14 & 23/08/27 & 10.2 & 23:37:45.74 & 14:42:37.1 & 250.9 & $<$30 & $<$2.2 \\
  SN 2014T & 14/02/22 & 23/08/25 & 9.5 & 14:36:04.98 & 02:20:34.2& 411.5 & $<$33 & $<$6.7 \\
  SN 2014Y & 14/03/02 & 23/07/30 & 9.4 & 07:23:33.38& 54:26:19.8& 172.0 & $<$28 & $<$0.99 \\
  SN 2014ab & 14/03/09 & 23/08/20 & 9.5 & 13:48:05.99  & 07:23:16.4& 101.2 & $<$40 & $<$0.50\\
  SN 2016iks & 16/11/20 & 23/08/27 & 6.8 & 23:07:21.53& 02:54:28.2& 282.7 & $<$26 & $<$2.5 \\
  SN 2016jae & 16/12/22 & 23/08/13 & 6.6 & 09:42:34.51& 10:59:35.4& 91.4 & $<$36 & $<$0.36 \\
  SN 2017eby & 17/04/01 & 23/08/25 & 6.4 & 14:38:50.74& 14:44:26.0& 368.1 & $<$30 & $<$4.9 \\
  SN 2017ifu & 17/10/28 & 23/08/28 & 5.8 & 02:50:55.03& $-$02:08:06.4& 887.8 & $<$24 & $<$23\\
  SN 2017hzw & 17/11/13 & 23/07/30 & 5.7 & 02:05:18.13& 53:12:13.4& 226.9 & $<$30 & $<$1.9 \\
  SN 2018cqj & 18/06/13 & 23/08/13 & 5.2 & 09:40:21.46& $-$06:59:19.8& 91.4 & $<$36 & $<$0.37 \\
  SN 2018gkx & 18/09/08 & 23/08/16 & 4.9 & 13:52:19.22 & 55:38:28.3& 646.0 & $<$33 & $<$17 \\
  SN 2019agi & 19/01/25 & 23/08/25 & 4.6 & 16:22:43.99&  24:01:09.7& 268.7 & $<$59 & $<$5.1 \\
  SN 2019rvb & 19/10/02 & 23/08/26 & 3.9 & 16:38:10.16& 68:27:49.1& 890.5 & $<$26 & $<$25 \\
  SN 2020kre & 20/05/21 & 23/08/20 & 3.3 & 13:10:29.87& $-$01:18:51.1& 640.9 & $<$77 & $<$38 \\
  SN 2020onv & 20/07/11 & 23/08/27 & 3.1 & 23:16:46.02& $-$23:18:37.1 & 435.8 & $<$38 & $<$8.6 \\
  SN 2020qxz & 20/08/08 & 23/08/16 & 3.0 & 18:04:00.23 & 74:00:50.3& 442.7 & $<$30 & $<$7.0 \\
  SN 2020abfe & 20/11/14 & 23/08/26 & 2.8 & 20:00:03.31 & 10:09:04.2& 426.1 & $<$42 & $<$9.2 \\ 
  SN 2020aeur & 20/12/22 & 23/08/13 & 2.6 & 09:26:20.39& 01:52:43.9& 226.9 & $<$37 & $<$2.3 \\ 
  SN 2020aekp & 20/12/27 & 23/08/26 & 2.7 & 15:43:11.39&  17:48:47.2& 236.1 & $<$49 & $<$3.3\\
  SN 2022erq & 22/03/11 & 23/08/16 & 1.4 & 18:33:25.36 & 44:05:11.7& 296.8 & $<$32 & $<$3.4 \\
  SN 2022esa & 22/03/12 & 23/08/26 & 1.5 & 16:53:57.60& $-$09:42:10.3& 100.2 & 1129 $\pm$ 52* & 13.6 $\pm$ 0.6*  \\
  SN 2022esa & 22/03/12 & 24/10/19 & 2.6 & 16:53:57.60 & $-$09:42:10.3& 100.2 & 799 $\pm$ 36* & 9.6 $\pm$ 0.4*\\
  \bottomrule
\end{tabular}
\tablecomments{* Measurements represent peak flux density for SN 2022esa using CASA \texttt{imfit} (see Section \ref{sec:obs} and \ref{sec:results})}
\label{tab:observations}
\end{table*}

Our targets were primarily observed for 10 minute durations (a few distant ones for 20 or 30 minutes) in C-band ($4-8$ GHz), using continuum mode, which covers the entire 4 GHz of bandwidth with 2 MHz-wide channels. Observations were obtained in A-configuration to achieve good localization accuracy (with restoring beams of $\sim$0.3\arcsec) and diminish the effects of diffuse emission from the SN host galaxies. Each target was observed between complex gain calibration observations lasting 1--2 minutes using calibrators that were selected for their brightness, high positional accuracy, and proximity to each respective target. Flux calibrations lasting $\sim$ 8 minutes were done at the beginning of each observing block using one of the following of the VLA's primary calibrators: 3C286, 3C138, 3C48, or 3C147.

We imaged and measured radio flux densities (or upper limits) for all of the targets in the sample using standard calibration and imaging procedures for radio continuum data followed in previous papers \citep[e.g.,][]{chomiuk16, Harris23, Hosseinzadeh2023}.
We used the standard VLA calibration pipeline implemented in \texttt{CASA} version 6.4.1.12 \citep{CASA2022} for reducing the continuum data from these observations. The pipeline applied online flags reported during observation, and additional flagging of RFI was done with the \texttt{rflag} algorithm. 
The pipeline performed iterative flagging--calibration cycles on the raw visibility data. 
The primary calibrators were used for delay, bandpass, and absolute flux density calibration, while the secondary calibrator was used for antenna-based complex gain calibrations. After the pipeline ran, each calibrated dataset was manually inspected and flagged for any remaining RFI.

Imaging was performed using the \texttt{tclean} task in \texttt{CASA}. We used \texttt{gridder=wproject} to mitigate the impact of non-coplanar baselines and sky curvature during wide-field imaging.%to project the $w$-dimension onto the ($u,v$) two-dimensional plane, imitating a coplanar array for the wide field of imaging. 
 For deconvolution, we used multi-term multi-frequency synthesis with \texttt{nterms=2} to account for frequency-dependent brightness in the imaging field. We applied Briggs weighting with \texttt{robust=0} to balance point source sensitivity with sidelobe contamination in our observations. We created images 6000 pixels wide, at a pixel scale of $0.07^{\prime\prime}$, which is approximately one-fifth of the synthesized beamwidth in VLA C-band and A-configuration. We cleaned our images for a maximum of 10$^4$ iterations or until the image root mean square contrast (RMS) was less than five times the peak residual  (\texttt{nsigma=5}). No self-calibration was necessary as our final images were close to the expected noise limit, with no artifacts from bright sources.

For each object in our sample, we recorded the pixel flux density and the 3$\sigma$ RMS noise at the location of the SN. Most of the SNe were non-detections, so we recorded 3$\sigma$ upper limits at 6 GHz, defined as the flux density at the SN location plus the 3$\sigma$ RMS noise in the 50-by-50 pixel region around the SN. If the flux density at the SN location was negative, we considered the 3$\sigma$ RMS noise as the upper limit. 

The majority of our sources turned out to be non-detections down to the 3$\sigma$ limits of $\lesssim$35 $\mu$Jy reported in Table \ref{tab:observations}. The implications of these non-detections regarding progenitor mass loss rates are discussed in Section \ref{sec:results}. One of our SNe, 2022esa, did produce detectable radio emission of $\sim$1.1 mJy, although as we discuss in Section \ref{sec:results}, the TNS classification of this object as a Ia-CSM is likely dubious. In order to confirm that the radio emission at the location of SN 2022esa is transient in nature, we obtained follow-up observations in the same frequency and array configuration as part of VLA/24B-381. The dataset was flagged, calibrated and imaged following the same procedure as the VLA/23A-328 datasets discussed above. We additionally recorded the peak flux density of the point source in both images using elliptical gaussian fitting with the \texttt{CASA} task \texttt{imfit}. Indeed, the flux density of the source decreased to 0.8 mJy after a year, corresponding to a 30\% decrease. These measurements are also presented in Table \ref{tab:observations}, and implications of the detections are further discussed in Section \ref{sec:results}.

%%__________________________________________
%%__________________________________________

\section{Radio Light Curve Model}
\label{sec:3}
In this section, we describe our model of synchrotron radio emission due to interaction with a wind-like CSM, which will be used to constrain the mass-loss rates of our SN Ia-CSM sample in Section \ref{sec:results}, given upper limits on their 6 GHz radio emission obtained in Section \ref{sec:obs}. 

\subsection{Synchrotron emission}
\label{subsec:radio_lcms}

We roughly follow the same synchrotron emission modeling procedure for Type Ia SNe that was used in previous papers \citep[e.g.,][]{chomiuk2012, chomiuk16, Harris23, Hosseinzadeh2023}, with a few modifications to the shock dynamics described in Section \ref{sec:randv}.\ We refer the reader to these papers for details of the model, but include a brief description here. 

The model follows the classical \cite{Chev82} model of synchrotron emission from SN ejecta interacting with CSM. The density profile of the CSM is assumed to be wind-like, with $\rho = \dot{M}/4\pi r^2 v_{wind}$, where $\dot{M}$ is the wind's constant mass-loss rate, and $v_{wind}$ is the wind velocity. Although idealized, a wind-like CSM is a reasonable approximation for single-degenerate progenitors with red-giant or AGB companions that are continuous mass-losing systems, and are considered strong candidates for our Ia-CSM SNe. Previous radio studies of SNe have often assumed a wind-like CSM since it provides convenient, self-similar solutions for shock dynamics and the resulting radio light curves, making it possible to explore vast parameter space of CSM properties \citep[see examples in][]{ChevFran2017}. Finally, a CSM wind profile is generally the model assumed in optical studies of Ia-CSM SNe for estimation of mass-loss rates (see discussion in Section \ref{sec:discussion}), so we assume the same for consistency with our radio-based results.

Broadly speaking, the optically-thin synchrotron luminosity depends on the volume of the shock, the energy density in the amplified magnetic field, and the energy density in relativistic electrons. This electron population is assumed to have a characteristic power-law spectrum, $N(E) = N_0 E^{-p}$, where we assume $p=3$ that is commonly observed in Type Ib/c SNe \citep{Soderberg2005, Soderberg2006} and is often assumed for Type Ia SNe \citep{chomiuk16}. The resulting optically-thin luminosity will have a frequency dependence $\propto \nu^{-1}$.
The energy density in relativistic electrons is assumed to be equal to $\epsilon_e \rho v^2_s$, where $\epsilon_e$ is the fraction of the shock energy in the relativistic electrons, $\rho$ is the density of CSM that the shock is interacting with at the time of radio observation, and $v_s$ is the shock velocity . The amplified magnetic field is assumed to scale with the shock energy as $B^2 = 8\pi \epsilon_B \rho v^2_s$ \citep{Chomiuk2009, Thompson2009}, where $\epsilon_B$, analogous to $\epsilon_e$, is the fraction of shock energy in the amplified magnetic field.  At early times, we assume this emission will be suppressed by synchrotron self-absorption and free-free absorption by the dense CSM. For synchrotron self-absorption, we use the form of the radio spectrum laid out in \cite{Chevalier1998} that self-consistently accounts for a $L_\nu \propto \nu^{\textbf{2.5}}$ turnover at lower frequencies, where the self-absorption optical depth $>$1. For free-free absorption, we multiply the luminosity by a factor $\epsilon^{-\tau_{\mathrm{ff}}}$, where $\tau_{\mathrm{ff}}$ is the free-free optical depth at a given wavelength ($\lambda$), which we derive from Equation (2.3) of \cite{FranClaes1996} as:
 \begin{equation} \label{eq:tauff} 
\tau_{\mathrm{ff}}\ \approx 0.46\,\lambda^2 \left(\frac{\dot{M}_{-5}}{v_{wind,10}}\right)^2 \ T_5^{-3/2} \ v_{s,4}^{-3} \left(\frac{t}{11.57 \mathrm{d}}\right)^{-3}
\end{equation}
where $\dot{M}_{-5}$ is the mass loss rate in units of $\mathrm{10^{-5}\ M_{\odot} \ yr^{-1}}$, ${v_{wind,10}}$ is the wind velocity in units of $\mathrm{10 \ km \ s^{-1}}$, $T_5$ is the wind temperature in $\mathrm{10^5 \ K}$, $v_{s,4}$ is the shock velocity in $\mathrm{10^4 \ km \ s^{-1}}$, and $t$ is the time elapsed between SN explosion and the radio observation in question. 
%We see curves corresponding to smaller rates of mass-loss at earlier times and curves with greater mass-loss rates at later times ,  therefore putting the curve produced by the smaller rate of mass-loss to the left (or before, chronologically) of the curve with the greater rate of mass loss. If these curves indeed contain the point of observed luminosity and time for the SN of interest, they will intersect at this point, therefore placing the point on the falling portion of the curve with the smaller mass-loss rate and on the rising portion of the curve with the greater rate of mass-loss. Therefore, in a Figure where the luminosity is plotted against time, we will find the 'lower limit' on the falling portion of the chronologically earlier curve,  and the 'upper limit' on the rising portion of the curve which started at a later time; this idea becomes clear when comparing \ref{fig:mlr_ro} with the values recorded in \ref{tab:tc17mlrs}.
%However, when plotting the luminosity against rates of mass-loss (i.e. Figure \ref{fig:2011dz}), the limits can be perceived intuitively, as they are shown in order of increasing value with no relation to time.

\begin{table}
  \begin{tabular}{l|lll}
  \toprule
   &Model \textit{A}&Model \textit{B}&Model \textit{C}\\
  \midrule
  \large 
  $\epsilon_{{B}}$&   0.1 &   0.1 &   0.01 \\
  $T_{{wind}}$ (K)&  10$^{5}$ &   2$\times$10$^{4}$ &   2$\times$10$^{4}$ \\
  \hline
  \large
  $\epsilon_{{e}}$&   0.1 &   0.1 &   0.1 
  \\
  ${v_{wind}}$ (km s$^{-1})$ & 100 & 100 & 100
  \\
  ${\nu}$ (GHz) & 6 & 6 & 6
  \\
  \bottomrule
  \end{tabular}
  \caption{Parameter values of the different radio light curve models for wind-like CSM explored in this paper.}
  %Description: This table presents the values of  $\epsilon_{b}$, T$_{wind}$ for Models A to C, aforementioned in Table \ref{tab:tc17mlrs}.
  \label{tab:params}
\end{table}

Our goal is to obtain constraints on the mass-loss rate given the observed radio luminosity (or upper limit) for each SN, but these constraints will depend on the values of several tunable parameters in our model. Based on \cite{Harris23}, the parameters expected to have the largest impact on the interpretation of the radio upper limits are $\epsilon_B$, $\epsilon_e$, and $T_{wind}$. As per convention, we assume the range of $\epsilon_B=0.01-0.1$ \citep{chomiuk2012}, while keeping $\epsilon_e=0.1$ for every model (see Section \ref{sec:discussion} for a discussion of these ranges). The values for $T_{wind}$ are expected to vary between $2\times10^4$ K to $10^5$ K for wind-like CSM around SNe \citep{Lundqvist2013}. In reality, both parameters will likely vary with time as the shock slows down and the post-shock conditions change, but for simplicity, we discuss our results for different constant values of $T_{{wind}}$ and $\epsilon_B$ (Table \ref{tab:params}) in order to provide an idea of the extent to which the various parameters affect the shape of the light curve, and therefore how they change the corresponding mass-loss rates. 

The synchrotron luminosity, corrected for absorption, is estimated using the equations detailed in 
\citet{Chevalier1998} and \citet{chomiuk16}, and the dynamics of the shock detailed in the next sub-section.  We follow the lead of \cite{Harris23} and \cite{Chev82} in using a volume filling factor of $f=0.5$ for these synchrotron models.
For the remainder of this paper, we will be assuming $v_{wind} = 100$ km s$^{-1}$ and a frequency  $\mathrm{\nu = 6 \ GHz}$ for the light curve calculations.

\subsection{Shock radius and velocity} \label{sec:randv}
\label{subsec:asymptotic}
\begin{figure*}
\hspace{-1.9cm}
\includegraphics[width=1.2\textwidth]{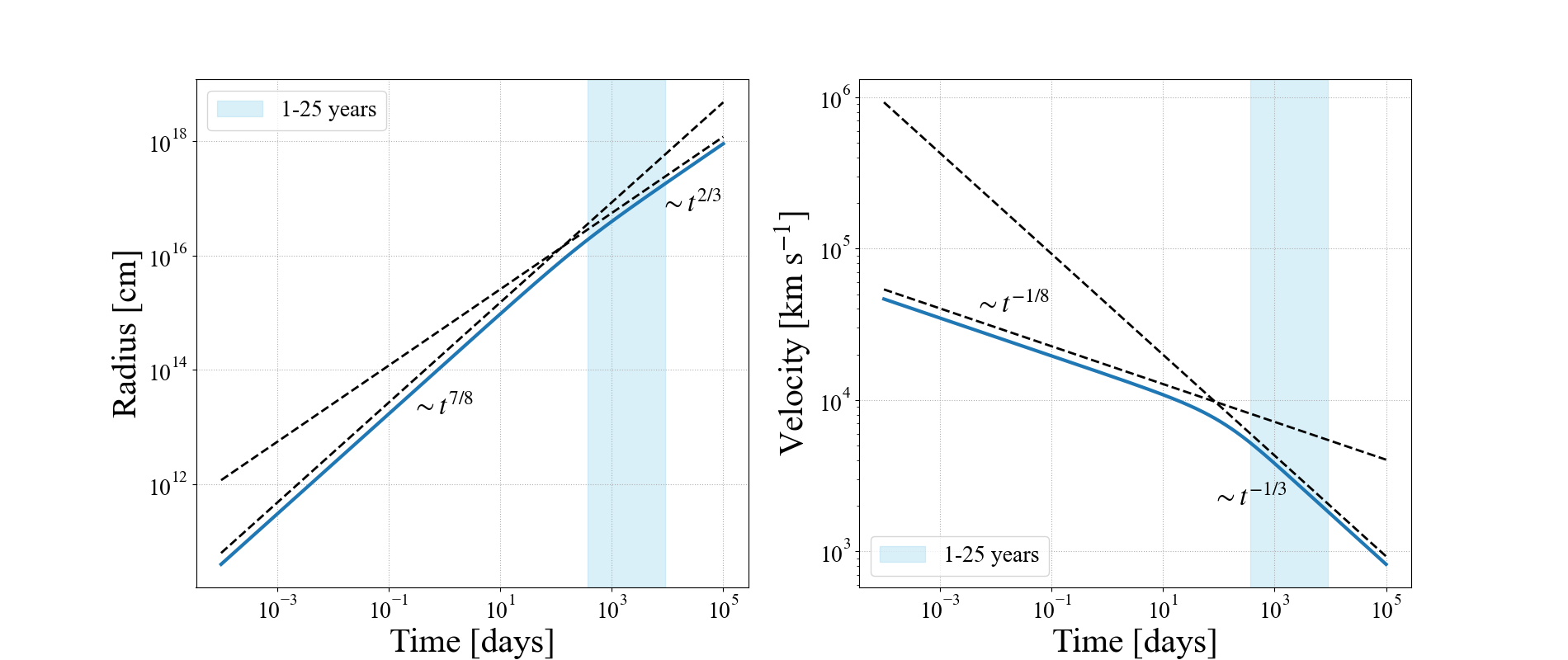}
  \caption{The time evolution of the forward shock radius (left panel) and velocity (right panel) from the TC17 model (Section \ref{sec:randv}) for the case of $E_{51}= 1.0$, $M_{ej}= 1.44$ \Msun, and $\dot{M}/v_{wind}$ = 0.1 \masslossunit. The vertical blue shaded section represents the time range at which our radio observations were obtained. Dashed lines represent the asymptotic solutions for the ejecta-dominated ($n=10$ density profile) and Sedov-Taylor cases.}
  \label{fig:fsr_vel}
\end{figure*}

The evolution of the radio light curves depends on the shock dynamics (i.e., the radius and velocity). For this paper, we use the set of solutions from \citet[][hereafter, TC17]{TC17} that describes the SN blast wave as it evolves from the ejecta-dominated (ED) phase to the Sedov-Taylor (ST) phase, in contrast with our previous papers (e.g., \citealt{chomiuk2012}, \citealt{chomiuk16}, etc.) where we assume the SN shock is ED. The relevant CSM density range for SNe Ia-CSM can be much higher than normal SNe Ia, so the swept-up CSM can be comparable to or even exceed the ejecta mass within the timescales of our observations. For example, for $\dot{M} \approx 1.0\times10^{-2}$ \Msunyr and $v_{wind} = \mathrm{100 \ km\ s^{-1}}$, Equation 25 in \cite{TC17} predicts that the swept-up mass will become dynamically important (i.e.,  $\mathrm{\geq  M_{ej}}$) as early as $\mathrm{7.9 \ yrs}$ after explosion. At this and later times, the radius and velocity is expected to follow a ST solution, and the TC17 model accounts for this transition by smoothly connecting the ED phase to the ST phase. From these solutions, we are able to determine a shock radius and velocity for ejecta interacting with wind density profiles at late observation times.

For the steep ($n>5$) ejecta profiles that are typically assumed for SNe, Eqs.\ (24) and (25) from \cite{TC17} give:
\begin{equation}
    R_s^*(t^*) = [(\zeta_{s}t^{*(n-3)/(n-s)})^{-\alpha} + ({\xi}t^{*2})^{-{\alpha}/(5-s)}]^{-1/\alpha},
\end{equation} \label{eq:rb1}
and
\begin{equation}
    v^*_b(t^*) = \frac{d R^*_s}{d t^*}, 
\end{equation} \label{eq:vb1}
respectively. Here $R_s^*$, $v_s^*$ and $t^*$ are dimensionless radius, velocity and age, which are related to the physical radius ($R_s$), velocity ($v_s$) and age ($t$) of the shock as $R^*_s = R_s/R_{ch}$, $v^*_s = v_s/v_{ch}$ and $t^*_s = t/t_{ch}$, where ``$X_{ch}$'' denote the characteristic variables that are defined as:
\begin{equation}\label{eq:r_ch}
    R_{ch} = (12.9\ \mathrm{pc})\ M_{ej}\ \dot{M}_{-5}^{-1}\ v_{wind,10}
\end{equation}
\begin{equation}\label{eq:t_ch}
    t_{ch} = (1.77\times10^3\ \mathrm{yr})\ E_{51}^{-1/2}\ M_{ej}^{3/2}\ \dot{M}_{-5}^{-1}\ v_{wind,10}
\end{equation}
\begin{equation}\label{eq:v_ch}
    v_{ch} = (7.12\times10^3\ \mathrm{km\ s^{-1}})\ E_{51}^{1/2}\ M_{ej}^{-1/2}.
\end{equation}
where 
%${\dot{M}_{-5}}$ is the wind mass-loss rate in units of $10^{-5}\ M_{\odot} \ \mathrm{yr^{-1}}$, 
${M_{ej}}$ is the ejecta mass in terms of ${M_{\odot}}$, 
%$v_{w,10}$ is the wind velocity in units of $\mathrm{10\ km\ s^{-1}}$, 
and ${E_{51}}$ is the kinetic energy  in units of $10^{51}$ erg. 

For a Type Ia-CSM SNe, we follow the model of an ejecta envelope with a steep density profile $n = 10$, and a wind density profile $s = 2$ \citep{Harris23}. The values of the coefficients $\zeta_s$, $\alpha$, and $\xi$ are drawn from Tables 4 and 6 of \cite{TC17} for an $n=10$ ejecta profile and wind-like ($s=2$) CSM profile. The resulting solution for the dimensionless radius and velocity, after some algebraic re-arrangement, is
\begin{equation}\label{eq:fsr}
    R_s^*(t^*) = 1.03t^{*(0.875)}[1+3.52t^{*(0.95)}]^{-0.219}
\end{equation}
and
\begin{equation}\label{eq:fsv}
    v^*_s(t^*) =  1.03 t^{*(-0.125)}[1+2.79t^{*(0.95)}][1+3.52t^{*(0.95)}]^{-1.219}.
\end{equation}
Throughout the paper, we will assume a standard SN Ia explosion with $M_{ej} =1.44\ M_{\odot}$ and $E_{51}=1$.

Figure \ref{fig:fsr_vel} illustrates the evolution of the shock radius and velocity from Eqs. \eqref{eq:fsr} and \eqref{eq:fsv}. The radius follows the self-similar driven wave (SSDW) solution (\citealt{Chev82}; \citealt{Nadez85}), with $R \boldsymbol{\propto}$ $t^{7/8}$  at early times (i.e. the ejecta-dominated phase) , and then approaches  $R \boldsymbol{\propto}$ $t^{2/3}$ for late times (i.e., the Sedov-Taylor phase, following the self-similar Sedov-Taylor solution) (\citealt{Taylor46}; \citealt{Sedov59}).  Since $v_s = dR/dt$, we see a trend of $v_s \boldsymbol{\propto} t^{-1/8}$ in the ED phase, which then approaches $v \boldsymbol{\propto} t^{-1/3}$ during the ST phase. The onset of the ST phase will occur earlier for denser CSM. 
%This behavior is as predicted in \cite{TC17} for both the forward shock radius and velocity over time, for our chosen values of $s=2$ and $n=10$. 
%In the old/previously assumed model, the mass-loss rate and the luminosity have a direct correlation. In the TC17 model, however, the luminosity increases with mass loss rate up until a certain point, at which it begins to decrease. The point at which the luminosity decreases is where we begin to see the influence of the Sedov Taylor phase, as discussed in \ref{subsec:radio_lcms}. 

%Comparing the two phases, we see that the ED phase has a much steeper slope than the ST phase, corresponding to a greater increase in radius over time; Therefore, we are seeing a greater rate of expansion during the ED phase; The shock then expands at a slower rate during the ST phase. As for the velocity, we know that a change in velocity gives acceleration; therefore we are seeing a deceleration of shock with the negative scaling relation. The ST phase has a much steeper decline than the ED phase, meaning it will have a greater rate of deceleration of the shock. All of these comparisons are can be visually determined from the plots in Figure \ref{fig:fsr_vel}.

%The implications of these values and the changes in scaling relations are further explained within subsection \ref{subsec:importance17tc}, which can be found in the Results (\ref{sec:results}. A comparison of the TC17 model with our old models is discussed in Section  \ref{app:A}.

\begin{figure*}
    \hspace*{-0.8cm}   \includegraphics[width=1.1\textwidth]{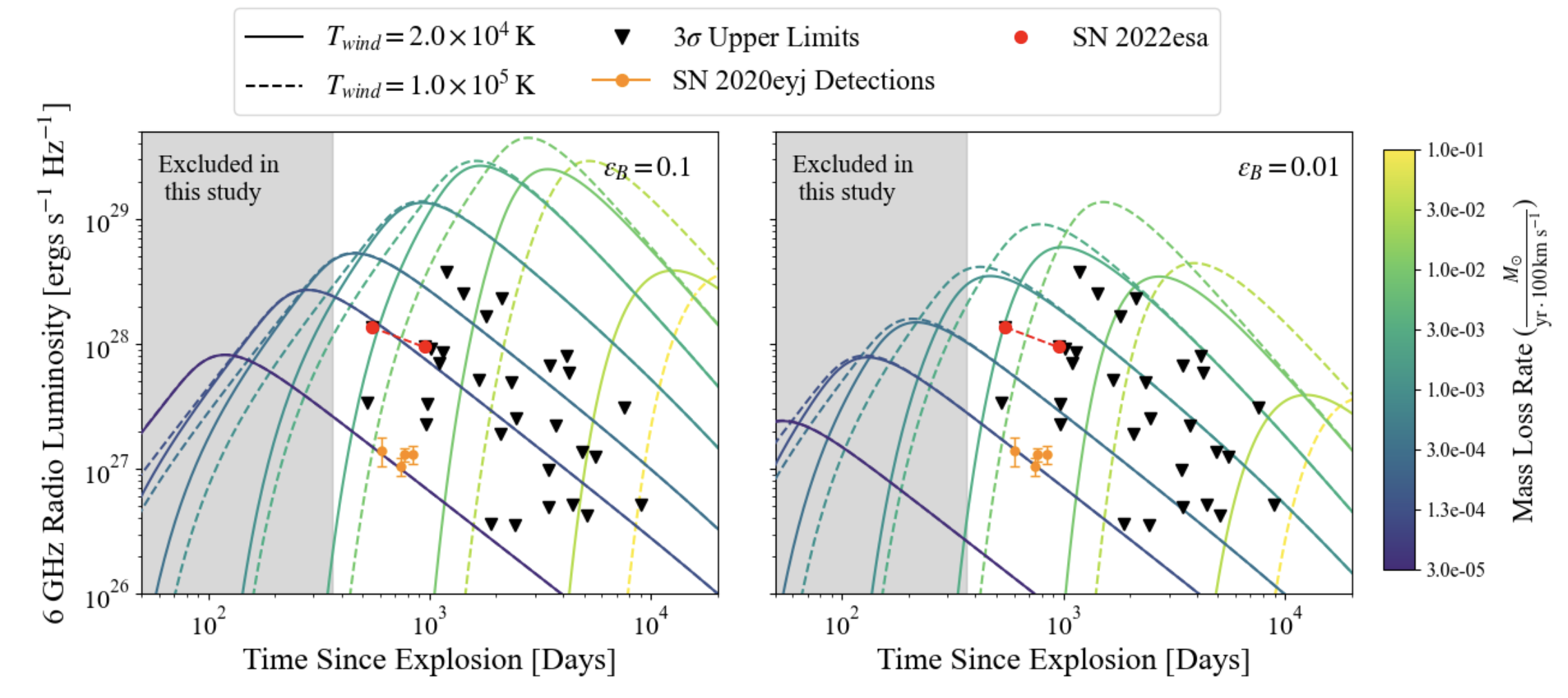}
    \caption{Predicted 6 GHz radio light curves for SNe Ia-CSM, shown alongside radio observations in this paper and from the literature. Solid curves correspond to the case of $T_{{wind}} = 2.0 \times$10$^{4}$ K, and dashed curves for $T_{{wind}} = $10$^{5}$ K. Left panel shows the case of $\epsilon_B = 0.1$, and the right panel shows $\epsilon_B = 0.01$ (both panels assume $\epsilon_e = 0.1$). Light curves span a range of $\dot{M}/v_{wind}$ from $3 \times 10^{-5}$ M$_{\odot}$ yr$^{-1}$ (dark blue lines) to $0.1$ M$_{\odot}$ yr$^{-1}$ (yellow lines), assuming $v_{wind} = 100$ km s$^{-1}$. The shaded gray region represents SNe Ia $<$ 1 yr, which are excluded in this study. SNe with 3$\sigma$ upper limits are plotted as inverted black triangles. The radio detections of SN 2022esa from 2023 and 2024 are shown as red circles connected by a dashed line. Literature measurements of SN 2020eyj from \cite{Kool23+} and \cite{yang23} (scaled to 6 GHz) are shown as orange points. }
    %See Section \ref{sec:results} for more details.}
    %. Several detections of this SN are plotted in orange; two of which are plotted with data from \cite{Kool23+}, where the data for the other two are from \cite{yang23}.  } 
\label{fig:summary_plot}
\end{figure*}

\subsection{Basic Properties of Radio Light Curves} \label{subsec:basic_properties}

The radio light curves from the model described in the previous sub-sections are summarized in Figure \ref{fig:summary_plot}. The shape of the light curves is similar to what has been seen in previous papers.
%\textbf{(For information on how the luminosity is calculated, see \citet{Chevalier1998} and \citet{chomiuk16}).} 
At early times absorption processes dominate, and the radio luminosity increases in proportion to the emitting area.
As the SN expands, the optical depths of the absorption processes decrease, and the light curves peak when $\tau \approx 1$, and then decline. This fading is primarily because both the energy in the electrons and the magnetic fields that are powering the emission are proportional to the shock energy ($\sim \rho v_s^2$). While the pre-shock density $\rho$ decreases as $r^{-2}$, the shock also continues to decelerate as it loses energy to interaction with the surrounding, resulting in an overall decrease in synchrotron emission with time. 

The different color lines in Figure \ref{fig:summary_plot} explore a sequence of $\dot{M}/v_{wind}$ values. As the CSM becomes denser (i.e., higher $\dot{M}/v_{wind}$), the light curves become brighter due to the increasing reservoir of shock energy ($\rho v_s^2$) available for the amplification of magnetic fields and relativistic electrons. The duration of the free-free absorbed phase of the light curve is also longer for denser CSM (see Eq\ \ref{eq:tauff}), leading to the light curves peaking later in time. This is why early-time radio observations would have missed detection of SNe with denser CSM. Above a certain mass-loss rate however ($\gtrsim 5 \times 10^{-3}$ \Msunyr), the light curves begin to become fainter. This is primarily because the onset of the ST phase occurs \emph{before} the light curve becomes optically-thin (i.e. $\tau_{\mathrm{ff}}=1$). The light curve in the ST phase declines faster because of the faster deceleration of the shock ($v \boldsymbol{\propto} t^{-1/3}$). On top of this, the age at which the light curve peaks is delayed during the ST phase as the shock is evolving slower, which delays the age at which $\tau_{\mathrm{ff}}=1$ (Eq. \ref{eq:tauff}). The combination of these effects produces light curves that peak and then decline at fainter luminosities for higher mass-loss rates. This contrasts with the behavior of light curves evolving with an ED-only solution, and this is discussed further in Appendix \ref{app:A}.

Figure \ref{fig:summary_plot} also illustrates the impact of the parameters $\epsilon_B$ and $T_{wind}$. Lower $\epsilon_B$ means a smaller fraction of $\rho v^2_s$ amplifies the magnetic fields, resulting in weaker fields, and thus fainter luminosities. We note that the parameter $\epsilon_e$, which we hold constant in this paper but can be uncertain, will have a similar magnitude of effect on the light curves as $\epsilon_B$. The $T_{wind}$ parameter affects the free-free optical depth $\tau_{\mathrm{ff}} \boldsymbol{\propto }T_{wind}^{-3/2}$ as seen in Eq. \eqref{eq:tauff}, so hotter CSM is less opaque to synchrotron emission. We therefore see the light curves for the hotter CSM (dashed lines in Figure \ref{fig:summary_plot}) become optically thin at earlier times. The difference between the $T_{wind}=1 \times 10^5$ K and $2 \times 10^4$ K becomes more dramatic at higher mass-loss rates, because of the effect of the ST phase described above. %Since $T_{wind}$ only affects the free-free optical depth, there is no difference in the light curves during the optically-thin phase for the two values of $T_{wind}$.

\subsection{Measuring mass-loss rates with the light curves} \label{sec:measuremassloss}

\begin{figure}
    \hspace*{-0.7cm}
     \includegraphics[width=1.2\linewidth]{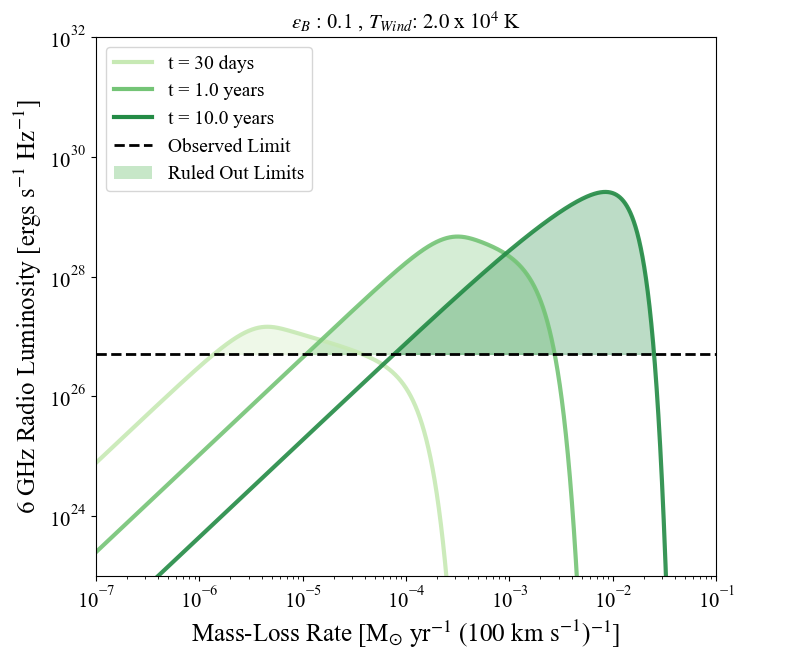}
    \caption{Illustration of how we use radio upper limits to constrain the range of \emph{ruled-out} mass-loss rates $\dot{M}/v_{wind}$ (green shade). Each light curve (green solid) shows the predicted luminosity for different mass-loss rates at a given SN age (30 days, 1 year, 10 years). The range of mass-loss rates where the predicted luminosity exceeds the measured upper limit (dashed line) are ruled out. Due to the light curve shape for wind-like CSM, later-time observations are more effective at ruling out a greater (and higher) range of mass-loss rates. }
    \label{fig:2011dz}
\end{figure}

We apply the light curve model described in Section \ref{subsec:radio_lcms} to the observed luminosities (or upper limits) from Table \ref{tab:observations} to constrain the possible values of $\dot{M}/v_{wind}$ for each SN, using the different parameter combinations describe in Table \ref{tab:params}. In the case of a detection, we simply check for what value of $\dot{M}/v_{wind}$ threads the light curve through the data point(s). For non-detections, which represent the majority of our sample, we can constrain the range of mass-loss rates ruled out by the luminosity limit as illustrated in Figure \ref{fig:2011dz}. At the given age of the SN, there will be a range of mass-loss rates for which the resulting luminosities at that age exceed the 3$\sigma$ upper limit. These are the shaded regions in Figure \ref{fig:2011dz}, and represent the range of $\dot{M}/v_{wind}$ that can be ruled out. The lower bound of this range comes from the optically-thin (decreasing) part of the light curves, while the upper bound comes from the optically-thick (rising) parts (this idea is better illustrated through the comparison of the variance in the optically-thick portion of the light curves in the leftmost panel of Figure \ref{fig:summary_plot}  to the upper limits derived for Models A and B in Table \ref{tab:tc17mlrs}). The range will additionally depend on the assumed values of $\epsilon_B$ and $T_{wind}$, and we discuss this further in Section \ref{sec:results}.

 Figure \ref{fig:2011dz} also nicely illustrates the benefit of obtaining late-time observations for our sample. For the same upper limit, observations at 30 days would have only ruled out an order-of-magnitude range of possible mass-loss rates, while at 10 years, more than a two orders-of-magnitude range is ruled out, shifted to the \emph{higher} values of $\dot{M}/v_{wind}$ that are more relevant for SNe Ia-CSM. This is again due to the free-free absorption described in the previous section, which suppresses the radio emission from SNe for longer periods at higher mass-loss rates.

\section{Results} \label{sec:results}

\begin{figure*}
    \centering
    \includegraphics[width=1\linewidth]{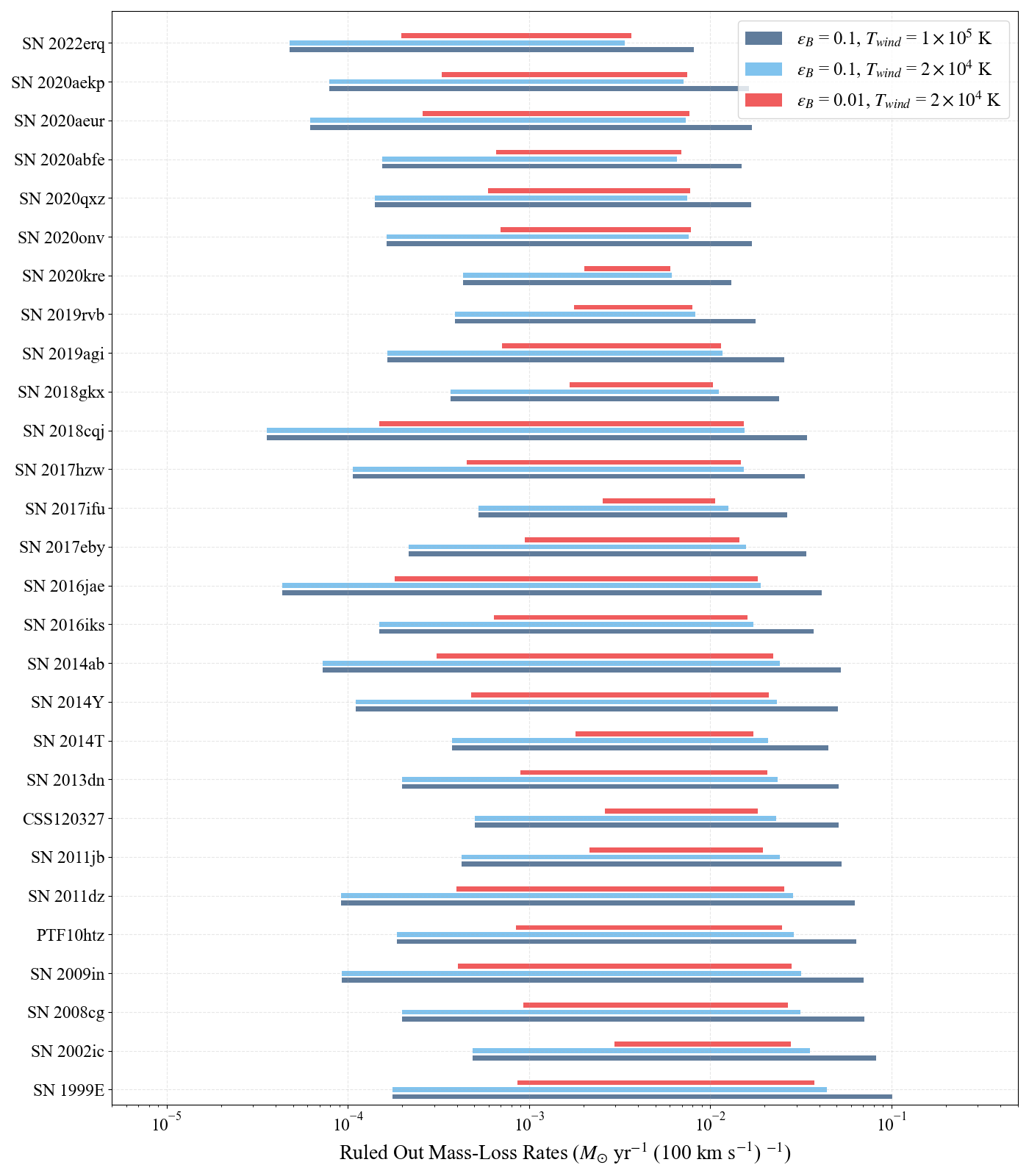}
    \caption{The range of mass-loss rates ruled out by the radio upper limits for each SN. For each SN, we show ranges for the three models in Table \ref{tab:params} for the parameters $\epsilon_B$ and $T_{wind}$. See Section \ref{sec:results} for details.}
    \label{fig:mlr_ro}
\end{figure*}

Most of the SNe (28 out of 29) in our sample are non-detections at 6 GHz, and their 3$\sigma$ upper limits are shown in Figure \ref{fig:summary_plot}. Based on the synchrotron emission model in Section \ref{sec:3}, these upper limits rule out a range of mass-loss rates for each SN, shown in Figure \ref{fig:mlr_ro}, with the specific ranges for each SN reported in Table \ref{tab:tc17mlrs}. %Assuming a wind velocity $v_{wind} = $  $\mathrm{100 \ km\ s^{-1}}$, we have found the average constrained mass-loss ranges calculated with the TC17 model to be between $(2-9)\times10^{-4}$ \masslossunit
%for the lower limits of each model, and $(1-4)\times10^{-2}$ \masslossunit for the upper limits. The actual calculated ranges can be found in Table \ref{tab:tc17mlrs}.

\begin{table*}
  \centering
  \caption {The range of ruled-out mass-loss rates for our Ia-CSM SNe computed with the \cite{TC17} model. For each model from Table \ref{tab:params}, we note the lower and upper limit of the ruled out range in \masslossunit. These ranges are shown visually for each SN in Figure \ref{fig:mlr_ro}.}
  \begin{tabular}{l||l|l|l}
  \toprule
   SN &Model \textit{A}&Model \textit{B}&Model \textit{C}\\
  \midrule
  SN 1999E & 1.8$\times$10$^{-4}$ - 1.0$\times$10$^{-1}$& 1.8$\times$10$^{-4}$ - 4.4$\times$10$^{-2}$ & 8.7$\times$10$^{-4}$ - 3.8$\times$10$^{-2}$\\
  SN 2002ic & 4.9$\times$10$^{-4}$ - 8.2$\times$10$^{-2}$& 4.9$\times$10$^{-4}$ - 3.6$\times$10$^{-2}$& 3.0$\times$10$^{-3}$ - 2.8$\times$10$^{-2}$\\
  SN 2008cg & 2.0$\times$10$^{-4}$ - 7.1$\times$10$^{-2}$& 2.0$\times$10$^{-4}$ - 3.2$\times$10$^{-2}$& 9.4$\times$10$^{-4}$ - 2.7$\times$10$^{-2}$\\
  SN 2009in & 9.3$\times$10$^{-5}$ - 7.0$\times$10$^{-2}$& 9.3$\times$10$^{-5}$ - 3.2$\times$10$^{-2}$& 4.1$\times$10$^{-4}$ - 2.8$\times$10$^{-2}$\\
  PTF10htz & 1.9$\times$10$^{-4}$ - 6.4$\times$10$^{-2}$& 1.9$\times$10$^{-4}$ - 2.9$\times$10$^{-2}$& 8.5$\times$10$^{-4}$ - 2.5$\times$10$^{-2}$\\
  SN 2011dz & 9.2$\times$10$^{-5}$ - 6.3$\times$10$^{-2}$& 9.2$\times$10$^{-5}$ - 2.9$\times$10$^{-2}$& 4.0$\times$10$^{-4}$ - 2.6$\times$10$^{-2}$\\
  SN 2011jb & 4.2$\times$10$^{-4}$ - 5.3$\times$10$^{-2}$& 4.2$\times$10$^{-4}$ - 2.4$\times$10$^{-2}$& 2.2$\times$10$^{-3}$ - 2.0$\times$10$^{-2}$\\
  CSS120327 & 5.0$\times$10$^{-4}$ - 5.1$\times$10$^{-2}$& 5.0$\times$10$^{-4}$ - 2.3$\times$10$^{-2}$& 2.6$\times$10$^{-3}$ - 1.8$\times$10$^{-2}$\\
  SN 2013dn & 2.0$\times$10$^{-4}$ - 5.1$\times$10$^{-2}$& 2.0$\times$10$^{-4}$ - 2.4$\times$10$^{-2}$& 9.0$\times$10$^{-4}$ - 2.1$\times$10$^{-2}$\\
  SN 2014T & 3.8$\times$10$^{-4}$ - 4.5$\times$10$^{-2}$& 3.8$\times$10$^{-4}$ - 2.1$\times$10$^{-2}$& 1.8$\times$10$^{-3}$ - 1.7$\times$10$^{-2}$\\
  SN 2014Y & 1.1$\times$10$^{-4}$ - 5.1$\times$10$^{-2}$& 1.1$\times$10$^{-4}$ - 2.3$\times$10$^{-2}$& 4.8$\times$10$^{-4}$ - 2.1$\times$10$^{-2}$\\
  SN 2014ab & 7.2$\times$10$^{-5}$ - 5.3$\times$10$^{-2}$& 7.2$\times$10$^{-5}$ - 2.4$\times$10$^{-2}$& 3.1$\times$10$^{-4}$ - 2.2$\times$10$^{-2}$\\
  SN 2016iks & 1.5$\times$10$^{-4}$ - 3.7$\times$10$^{-2}$& 1.5$\times$10$^{-4}$ - 1.7$\times$10$^{-2}$& 6.4$\times$10$^{-4}$ - 1.6$\times$10$^{-2}$\\
  SN 2016jae & 4.4$\times$10$^{-5}$ - 4.1$\times$10$^{-2}$& 4.4$\times$10$^{-5}$ - 1.9$\times$10$^{-2}$& 1.8$\times$10$^{-4}$ - 1.8$\times$10$^{-2}$\\
  SN 2017eby & 2.2$\times$10$^{-4}$ - 3.4$\times$10$^{-2}$& 2.2$\times$10$^{-4}$ - 1.6$\times$10$^{-2}$& 9.5$\times$10$^{-4}$ - 1.5$\times$10$^{-2}$\\
  SN 2017ifu & 5.3$\times$10$^{-4}$ - 2.7$\times$10$^{-2}$& 5.3$\times$10$^{-4}$ - 1.3$\times$10$^{-2}$& 2.6$\times$10$^{-3}$ - 1.1$\times$10$^{-2}$\\
  SN 2017hzw & 1.1$\times$10$^{-4}$ - 3.3$\times$10$^{-2}$& 1.1$\times$10$^{-4}$ - 1.5$\times$10$^{-2}$& 4.6$\times$10$^{-4}$ - 1.5$\times$10$^{-2}$\\
  SN 2018cqj & 3.6$\times$10$^{-5}$ - 3.4$\times$10$^{-2}$& 3.6$\times$10$^{-5}$ - 1.6$\times$10$^{-2}$& 1.5$\times$10$^{-4}$ - 1.5$\times$10$^{-2}$\\
  SN 2018gkx & 3.7$\times$10$^{-4}$ - 2.4$\times$10$^{-2}$& 3.7$\times$10$^{-4}$ - 1.1$\times$10$^{-2}$& 1.7$\times$10$^{-3}$ - 1.0$\times$10$^{-2}$\\
  SN 2019agi & 1.7$\times$10$^{-4}$ - 2.6$\times$10$^{-2}$& 1.7$\times$10$^{-4}$ - 1.2$\times$10$^{-2}$& 7.1$\times$10$^{-4}$ - 1.2$\times$10$^{-2}$\\
  SN 2019rvb & 3.9$\times$10$^{-4}$ - 1.8$\times$10$^{-2}$& 3.9$\times$10$^{-4}$ - 8.3$\times$10$^{-3}$& 1.8$\times$10$^{-3}$ - 8.0$\times$10$^{-3}$\\
  SN 2020kre & 4.3$\times$10$^{-4}$ - 1.3$\times$10$^{-2}$& 4.3$\times$10$^{-4}$ - 6.2$\times$10$^{-3}$& 2.0$\times$10$^{-3}$ - 6.0$\times$10$^{-3}$\\
  SN 2020onv & 1.6$\times$10$^{-4}$ - 1.7$\times$10$^{-2}$& 1.6$\times$10$^{-4}$ - 7.6$\times$10$^{-3}$& 7.0$\times$10$^{-4}$ - 7.9$\times$10$^{-3}$\\
  SN 2020qxz & 1.4$\times$10$^{-4}$ - 1.7$\times$10$^{-2}$& 1.4$\times$10$^{-4}$ - 7.5$\times$10$^{-3}$& 5.9$\times$10$^{-4}$ - 7.8$\times$10$^{-3}$\\
  SN 2020abfe & 1.6$\times$10$^{-4}$ - 1.5$\times$10$^{-2}$& 1.6$\times$10$^{-4}$ - 6.6$\times$10$^{-3}$& 6.6$\times$10$^{-4}$ - 6.9$\times$10$^{-3}$\\
  SN 2020aeur & 6.2$\times$10$^{-5}$ - 1.7$\times$10$^{-2}$& 6.2$\times$10$^{-5}$ - 7.3$\times$10$^{-3}$& 2.6$\times$10$^{-4}$ - 7.7$\times$10$^{-3}$\\
  SN 2020aekp & 7.9$\times$10$^{-5}$ - 1.6$\times$10$^{-2}$& 7.9$\times$10$^{-5}$ - 7.1$\times$10$^{-3}$& 3.3$\times$10$^{-4}$ - 7.5$\times$10$^{-3}$\\
  SN 2022erq & 4.8$\times$10$^{-5}$ - 8.1$\times$10$^{-3}$& 4.8$\times$10$^{-5}$ - 3.4$\times$10$^{-3}$& 2.0$\times$10$^{-4}$ - 3.7$\times$10$^{-3}$\\
  %SN 2022esa & 5.73$\times$10$^{-5}$ - 7.86$\times$10$^{-3}$& 5.73$\times$10$^{-5}$ - 3.30$\times$10$^{-3}$& 2.38$\times$10$^{-4}$ - 3.61$\times$10$^{-3}$\\
  \bottomrule
  \end{tabular}
  \label{tab:tc17mlrs}
\end{table*}

We find that the radio upper limits rule out the mass-loss rate range between $\sim$$10^{-4} - 10^{-2}$ \Msunyr (assuming $v_w$=100 km s$^{-1}$) for most of our sample, with the exact ruled-out range varying with the SN age when the radio observation was carried out, as well as the assumed values of $\epsilon_B$ and $T_{wind}$. For example, decreasing $\epsilon_B$ by $\mathrm{1\ dex}$ (i.e. from $0.1$ to $0.01$) yields an increase of  $\mathrm{0.68\ dex}$  in the $\dot{M}$ lower limits and a decrease of $\mathrm{0.05 \ dex}$ in the upper limits, on average. The variation of $T_{wind}$ has more of an effect on the upper limits of $\dot{M}$, while there is a negligible effect on the lower limits when retaining the same value for $\epsilon_B$. For instance, when $\epsilon_B = 0.1$ stays the same and $T_{wind}$ increases by $\mathrm{0.7 \ dex}$ (from $\mathrm{2.0\times10^4 \ K}$ to $\mathrm{1.0\times10^5 \ K}$), the change in the lower limit is negligible while the upper limits increase by an average of $\mathrm{0.35 \ dex}$. Lowering the value of $\epsilon_B$ for the same $T_{wind}$ pushes down the light curve to fainter luminosities, which makes the lower limit of the ruled-out mass-loss rates less constraining. %, because the rate of rise of the free-free absorbed part of the light curves is faster than the rate of decline of the optically-thin part (Figure \ref{fig:summary_plot}). 
 %In contrast, lowering $T_{wind}$ for the same $\epsilon_B$ primarily affects the upper limit, as $T_{wind}$ only affects the free-free optical depth, and thus only the rising part of the light curve which constrains the upper limit (as discussed in Section \ref{subsec:basic_properties}). 

The range of excluded mass-loss rates progressively shifts higher for older SNe in Figure \ref{fig:mlr_ro}. For instance, we see that our radio observations of SN 2022erq rule out  $2.0\times10^{-4} < \dot{M} < 3.7\times10^{-3}$ M$_{\odot}$ yr$^{-1}$, for $\mathrm{\epsilon_B = 0.01, T_{wind} = 2\times10^4 \ K}$, and $v_{wind} = 100$ km s$^{-1}$, whereas we exclude $8.7\times10^{-4} < \dot{M} < 3.8\times10^{-2}$ M$_{\odot}$ yr$^{-1}$ for SN 1999E  for the same values of $\epsilon_B$, $T_{wind}$, and $v_{wind}$
(exact values are recorded in Table \ref{tab:tc17mlrs}). This is because of the effect observed in Figure \ref{fig:2011dz} and discussed in Section \ref{sec:measuremassloss}, where at older ages, the SN observations are more sensitive to emission from higher mass-loss rates that emerge progressively later in time. The progression in Figure \ref{fig:barh compare} is however non-monotonic, as some SNe are more distant than others, thus having shallower limits (e.g. the ruled out range of SN 2019agi is significantly smaller than the range for SN 2018cqj, even though they are a year apart, because SN 2018cqj is nearly three times closer). 

We briefly discuss SN 2022esa, the only SN radio-detected SN our sample. While it was initially classified as a SN Ia-CSM\footnote{\href{https://www.wis-tns.org/object/2022esa/classification-cert.}{TNS classification}}, re-inspection of the public TNS spectrum shows that the narrow lines are likely from the host galaxy, given the visible [\ion{O}{3}] and [\ion{S}{2}] lines in addition to the narrow Balmer lines. Removing these lines using {\tt SNID emclip} \citep{blondin2007} does not provide a very reliable classification. In addition, a subsequent spectrum obtained at the Nordic Optical Telescope on 2022-07-17 show better matches with a stripped envelope core collapse supernova (Type Ib/c), and late nebular spectra reveal lines of oxygen and carbon but no lines from iron-peak elements (YuJing Qin, private communication) which further argues for a core-collapse origin (Qin et al. in prep). A more detailed investigation of the object will be presented in Qin et al. in prep, but here we briefly note the radio evolution observed for 2022esa. Figure \ref{fig:lccs} shows the observed VLA measurements, showing a clear decline in radio emission similar to SN2020eyj, except that it much brighter. The slope of the decline is slightly shallower than our Sedov-phase light curves. The mass-loss rates that best-fit these data-points are in the range of $\left(1.2-1.5\right)\times10^{-4}$ \Msunyr ($\epsilon_B=0.1$) or $\left(5-6.4\right)\times10^{-4}$ \Msunyr ($\epsilon_B=0.01$).

We also re-interpret the radio data for SN 2020eyj, which is still the only radio-detected Ia-CSM, with the modified radio light curve model in this paper. SN 2020eyj was observed at 5.1 GHz with e-MERLIN in \cite{Kool23+} on 2021 November 19, then again during six consecutive days between 2022 April 6--12. This SN was also observed with the VLA in \cite{yang23} on both April 14th and June 26th of 2022. The data collected from these e-MERLIN and VLA observations are shown as orange points in Figure \ref{fig:summary_plot}. SN 2020eyj was consistent with constant or slightly decreasing flux over these radio observations, meaning we can assume the emission is optically thin. We scale the e-MERLIN measurements to 6 GHz assuming the luminosity $L_{\nu} \propto \nu^{-1}$. We find that the luminosities are consistent with mass-loss rates of $3.4 \times 10^{-5}$ \Msunyr for $\epsilon_B=0.1$ and $1.4 \times 10^{-4}$ \Msunyr for $\epsilon_B=0.01$. The assumed value of $T_{wind}$ should not affect the mass-loss rates, as the emission is optically-thin. Our measurements are lower than in \cite{Kool23+} because of their lower fitted values of $\epsilon_B$, which we return to in Section \ref{sec:discussion}.

\section{Discussion} \label{sec:discussion}

Our results indicate that the majority of SNe Ia-CSM do not produce luminous radio emission at 6 GHz on $\sim$1--25 yr timescales, 
%at least with flux densities brighter than $\sim$35 $\mu$Jy, 
in contrast with SN 2020eyj\footnote{We also note that 2020eyj was a He-rich Ia-CSM \citep{Kool23+}, which contrasts with our sample that are H-rich Ia-CSM.}.\ %, shows no visible late-time (1.5--25 yr after explosion) radio emission at 6 GHz similar to SN 2020eyj, indicating that the probability of finding such late-time radio emission is low, at least to the depths we have explored.
The non-detections of our sample also stand in contrast with Type IIn SNe, the core-collapse cousins of SNe Ia-CSM, where at least 10\% of the population with radio observations have detectable light curves that peak on timescales of about 2--34 years (substantially longer than other core-collapse SN subtypes) at luminosities comparable to our upper limits \citep[see][and references therein]{Chandra2018, Bietenholz2021}. One possibility is that, assuming both types of SNe explode with the canonical 10$^{51}$ erg energy, the Ia-CSM SNe on average have lower CSM densities than Type IIn SNe, and thus the radio-detectable Ia-CSM population may be peaking at times similar to SN\,2020eyj, i.e. a year or two after explosion (as opposed to longer few-decade timescales like SNe IIn). %are most likely to be detectable in the radio within a year or two of explosion at the depths probed by our VLA survey. 
A similar reasoning was applied to X-ray observations of SNe Ia-CSM, where it was argued that SN 2012ca remains the only X-ray-detected SN Ia because it was likely observed at an opportune moment, when the ambient medium was dense enough to produce visible X-rays but not too dense to suffer complete absorption \citep{Bochenek2018, Dwarkadas2023, Dwarkadas2024}. 
In this paper, we present only two SNe with radio observations at $\lesssim$2 yr ages -- SN 2022erq which was a non-detection, and SN 2022esa, which was likely a peculiar Ic. \cite{chomiuk16} compiles radio upper limits for six more SNe Ia-CSM in the first year of explosion. These observations imply that luminous radio emission in the first 1--2 years is also unlikely to be common, but more sensitive constraints and larger samples can further test this hypothesis in the future.
%Following up a larger sample of SNe Ia-CSM on these 1--2 year timescales in future will help test this hypothesis.

The range of mass-loss rates ruled out by our radio survey, roughly between $10^{-4}-10^{-2}$ \Msunyr, is somewhat in tension with the mass-loss rates derived from optical observations of SNe Ia-CSM. These rates are most commonly derived by modeling the broad component of the H$\alpha$ line as $L_{H\alpha}$ $\textbf{=}\mathbf{\frac{1}{4}}$ $\epsilon_{Ha} (\dot{M}/v_{wind})v_s^3$, assuming hydrogen in a spherically-symmetric CSM with a $r^{-2}$ profile is being collisionally excited by the shock \citep[e.g.,][]{Chugai1991, Salamanca1998}. The observed luminosities and velocities of $\sim$few $\times$ 10$^3$ km s$^{-1}$, measured from the broad H$\alpha$ line profiles  in the first $\sim$year following explosion, yield mass-loss rates typically in the range of $10^{-4}-10^{-2}$ \Msunyr, assuming $100\ \mathrm{km\ s^{-1}}$ winds and an efficiency $\epsilon_{H\alpha}=0.1$ \citep[e.g.][]{Aldering2006, Silverman2013, Sharma2023}.
This range of  mass-loss rates is generally inconsistent with our radio-derived constraints on $\dot{M}$ in Figure \ref{fig:mlr_ro} for the same CSM geometry. Based on the radio measurements alone, the required mass-loss rates would have to be lower than 10$^{-4}$ \Msunyr or higher than $10^{-2}-10^{-1}$ \Msunyr. However, we point out the radio observations presented here, obtained years after explosion, probe the CSM at significantly larger radii than the H$\alpha$ measurements. Specifically, optical H$\alpha$ implies mass-loss rates of $\sim 10^{-4}-10^{-2}$ \Msunyr at $R_{in}$ = $10^{15}-10^{16}$ cm, while our radio limits apply to radii in the range of 10$^{17}$-10$^{18}$ cm. Radio observations of several SNe Ia-CSM do exist in epochs overlapping H$\alpha$ measurements, but these also yield non-detections \citep{chomiuk16}.
In fact, these non-detections from earlier-time ($<$1 yr) radio observations  push the lower end of the ruled-out $\dot{M}$ range further down to $\lesssim$10$^{-5.5}$ \Msunyr ($\mathrm{100\ km\ s^{-1}}$)$^{-1}$ \citep[e.g.][]{Dilday12, chomiuk16}. 
While such low mass-loss rates are more consistent with realistic progenitors like symbiotic binaries \citep{Seaquist1990, chomiuk2012}, they may under-predict the H$\alpha$ luminosities, requiring a substantially more efficient $\epsilon_{Ha}$ or faster shock velocities than measured from the broadened H$\alpha$. 

Mass-loss rates higher than 0.1 \Msunyr $\mathrm{(100\ km\ s^{-1}})^{-1}$ are also allowed by the radio limits, and we do note that mass-loss rates derived from modeling the bolometric light curves of SNe Ia-CSM in \cite{Silverman2013} are $\gtrsim$$10^{-1}$ \Msunyr. However, these high values are more inconsistent with wind-like CSM in single-degenerate systems. In fact, such copious winds are extreme even for typical supergiant stars, and only observed in luminous blue variable eruptions \citep{Smith2014}. %; however, this could become more consistent with common-envelope like CSM.

A possible way to reconcile the tension between radio and optical observations of SNe Ia-CSM is if microparameters \epsb and \epse are significantly lower than the range assumed here ($10^{-2}-10^{-1}$). Lowering these parameters makes the radio luminosities fainter (as discussed in Section \ref{subsec:basic_properties}), which shrinks the range of ruled-out mass-loss rates further (Figure \ref{fig:mlr_ro}). For \epsb $<10^{-4}$, the light curves for the majority of our sample fall below our observed limits, and no mass-loss rate can be ruled out. \cite{Kool23+} showed that the bolometric light curve of SN 2020eyj suggested wind mass-loss rates of $10^{-4}$ to $3\times 10^{-3}$ \Msunyr for $v_{wind}=100$ km s$^{-1}$, which could fit the radio observations only if $\epsilon_B = 1.7 \times 10^{-3}$ or $1.5 \times 10^{-5}$. On the other hand, $\epsilon_B=0.1$ was consistent with the radio light curve of SN 2020eyj for a shell-like CSM geometry like in \cite{Harris2016} \citep[Figure 4 in][]{Kool23+}. Unfortunately, the observational and theoretical basis for determining values of $\epsilon_e$ and $\epsilon_B$ is still a topic of debate. Values of \epsb=0.01--0.1 have been inferred in relativistic SNe \citep[e.g.][]{ChevalierFransson2006, Soderberg2012}, and given the similar polytropic structure of their progenitors and white dwarfs, this range has been conventionally assumed for SNe Ia \citep{chomiuk2012}. Values of \epse and \epsb $\gtrsim$10$^{-3}$ have been inferred in shocks with velocities of a few $\times$ 10$^3$ km  s$^{-1}$ in the Kepler SN remnant \citep{Reynolds2021}, also a candidate for a Type Ia that exploded in a dense CSM \citep[e.g.,][]{Katsuda2015}. Similar values were also derived from radio light curves of some Type II SNe \citep{Chevalier2006}. Theoretically, a clear prescription of how these parameters change with the shock conditions and shock structure in collisionless shocks is also yet to be established \citep[see review in][]{Marcowith2016}, so we do not have firm evidence leading us to prefer lower values of $\epsilon_B$.

Another possible way of addressing the tension is if the CSM is not distributed as a wind, but rather discrete, dense shells produced by episodic mass-loss events such as common envelope ejections \citep[e.g.,][]{Livio2003} or swept-up nova-shells \citep[e.g.][]{WoodVasey2006, Moore2012}. Such shell-like geometries, with thicknesses of up to four times the inner radii, have been proposed in some SNe Ia showing delayed interactions, like PTF11kx and SN 2002ic \citep[e.g.][]{Dilday12, Graham2017, Harris2018}. This shock+shell interaction beginning a few months after explosion would power H$\alpha$, but then rapidly diminish once the shock crosses over the shell, eluding later radio observations. In fact, radio observations of SN 2020eyj were reproduced with a best-fit shell $\sim$0.3 \Msun located within 10$^{17}$cm, but the resulting luminosity is expected to diminish well below 10$^{27}$ \lumunit (the lower end of our limits in Figure \ref{fig:summary_plot}) by 10$^3$ days according to Figure 4 in \cite{Kool23+}. 
%Modeling such discontinuous or shell-like CSM \citep[e.g. with the formalism of][]{Harris2016, Harris2021} will be attempted in future papers, and will benefit from coordinated optical and radio observations to correctly constrain the extents of the shells.  

\begin{figure*}
    \centering
    \includegraphics[width=1\linewidth]{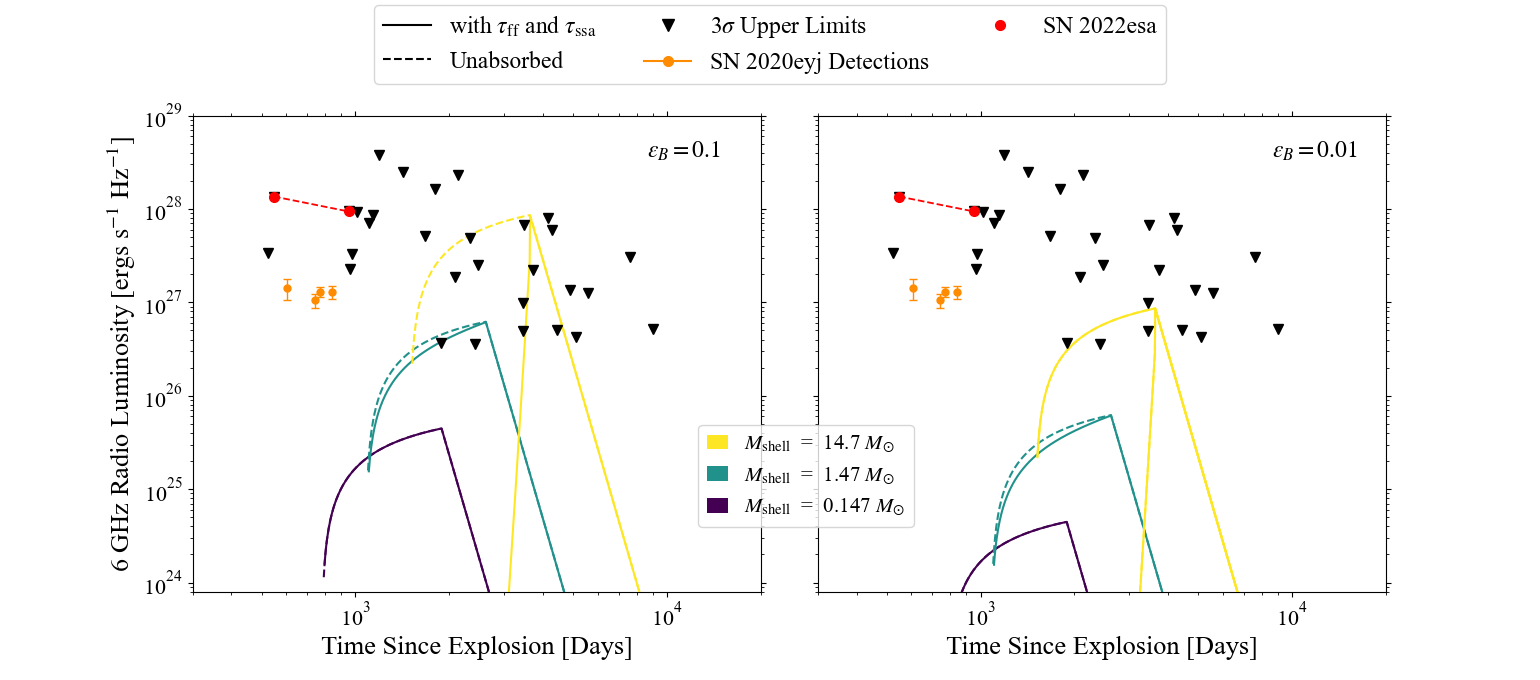} \caption{ 6 GHz light curves produced by interaction with shell-like CSM, compared with our radio observations of SNe Ia-CSM. The light curves accounting for absorption are solid, whereas the ones with no absorption are dashed.  
    The left panel shows models for $\epsilon_B = 0.1$, and the right panel for $\epsilon_B = 0.01$ (both panels use $\epsilon_e = 0.1$). 
    Each light curve's color is specific to the density and mass of the shell, with the dark purple color representing $\rho_{csm} = 10^{-20}$ g cm$^{-3}$ and $M_{shell} = 0.147 $ $ M_{\odot}$, teal corresponding to $\rho_{csm} = 10^{-19}$ g cm$^{-3}$ and $M_{shell} = 1.47 $ $ M_{\odot}$, and yellow depicting $\rho_{csm} = 10^{-18}$ g cm$^{-3}$ and $M_{shell} = 14.7 $ $ M_{\odot}$. 
    We assume $R_{in} = 10^{17}$ cm  and $f_{R} = 1$ for this model (see Section \ref{sec:discussion}). 
    %These light curves are overlaid with the data points from Figure \ref{fig:summary_plot} for comparison.
    }
    \label{fig:shellmodel}
\end{figure*}

%It is also possible that the CSM is not continuous and wind-like, but instead configured in discrete shells, as might be expected if the progenitor system lost mass impulsively. 
We consider the models of SN interaction with shell-like CSM from \citet{Harris2016,Harris2021}, where a shell is characterized by a constant mass density $\rho_{csm}$, an inner radius $R_{in}$, and a width of the shell ($f_R$, which is described as a fraction of $R_{in}$). Because shells are characterized by more parameters than wind-like CSM, detailed constraints on shell-like CSM are outside the scope of this paper. However, to give a sense of what sorts of shells can be constrained by current observations, Figure \ref{fig:shellmodel} shows a few example light curves for shells spanning two orders of magnitude in mass. Models assuming \epsb=0.1 are shown in the left panel and \epsb=0.01 in the right panel; in both panels, we take \epse = 0.1 and assume a temperature of the shell (which is assumed to be ionized) of ${2\times10^{4}}$ $\mathrm{ K}$. 
The models shown take $R_{in} = 10^{17}$ cm to yield interaction around the time of our observational constraints (smaller $R_{in}$ leads to earlier radio emission), and assume $f_R = 1$ (thicker shells lead to longer-lasting radio emission). The unabsorbed light curves are plotted as dashed lines, and rise rapidly after the start of interaction with the shell, and plummet once the shock traverses the shell. The unabsorbed peak spectral radio luminosity (in erg s$^{-1}$ Hz$^{-1}$) is: 
$$L_{\nu} = 1.7\times10^{54}\, \epsilon_e^2\, \epsilon_B\, \nu^{-1}\, \rho_{csm}^{8/7}\, r_{in}^{3/7}\, \left(1-(1+f_r)^{-1.28}\right)$$
Using the expressions of \citet{Harris2021}, we correct the radio light curves for both synchrotron self-absorption and free-free absorption, and plot these as solid lines. In the case of the densest shell ($\rho_{csm} = 10^{-18}$ g cm$^{-3}$; $M_{shell} =$ 14.7 M$_{\odot}$), absorption strongly affects the light curve, narrowing it to a brief burst emitted when the shock reaches the outer edge of the shell. The 1.47 M$_{\odot}$ shell model is barely affected by absorption, and the 0.147 M$_{\odot}$ shell model not at all.

Figure \ref{fig:shellmodel} demonstrates that for a shell to be detectable by our radio observations of SNe Ia-CSM, it needs to be quite dense and massive. The lower mass shells of $\sim$0.1 M$_{\odot}$ (which might be expected from a shell produced by several recurrences of a nova) are not expected to be detectable at radio wavelengths at these distances.
Our most sensitive limits can constrain shells of $\sim$1 M$_{\odot}$ (as might be expected from a common envelope ejection event). 
Our less sensitive limits require densities/masses an order of magnitude higher.
 At these higher densities, absorption is important, narrowing the light curve and requiring an observer to get very lucky to capture the transient radio signal. 
%We note that the radio observations of SN\,2020eyj were modelled by \citet{Kool23+} as a shell of mass 

And yet, the observations of SNe Ia-CSM presented here still have potential to rule out common envelopes surrounding the explosion. A significant fraction of planetary nebulae are the result of common envelope ejection, and systems like the Ring Nebula and NGC 6818 show densities of several $\times 10^2-10^3$ cm$^{-3}$ over radii spanning $10^{16}$ cm to several $\times 10^{17}$ cm \citep{Benetti03,ODell07}. These shells are so radially extended that they are not well modeled by the equations of \citet{Harris2016,Harris2021}, which are calibrated for $f_R = 0.1-1$, and they might be better approximated with a uniform medium. Interaction with a uniform medium of density $10^3$ cm$^{-3}$ yields a detectable radio luminosity, $3 \times 10^{27}$ erg s$^{-1}$ Hz$^{-1}$ on day 1000 for \epsb=0.1 (and becomes more luminous with time; \citealt{chomiuk16}). While more work is needed to predict the radio light curves expected from interaction with a planetary nebula-like CSM (especially when accounting for likely departures from spherical symmetry), this back-of-the-envelope estimate implies they are likely detectable around SNe Ia-CSM with current radio telescopes. The assumption of spherical steady mass loss is likely violated in most progenitors, but the limits we estimate in this paper still exclude average mass-loss densities in the ruled-out range.

%((we see the same upper limits for Twind 2e4 and 1e5, but their lower limits vary... CHECK THIS WITH SUMMARY PLOT.))((this is confirmed WITH WHAT PLOT OR CALCULATION, and expected Twind has very little sway on the optically thin portion of the light curve, as there is no longer any free-free absorption happening. The rising portion of the curve, however, is strongly affected by a change in Twind, as the value of Twind impacts the free-free absoprtion, as seen in the tauff function from our fiducial luminosity model (insert tauff code here.)  ))

%((caveats:... optical and radio done at two different times.
%CSM might not be a steady wind,, no fundamental reason. 
%geometry, physics.))))

\section{Conclusions}\label{sec:6}
Motivated by the recent (and first ever) radio detection of SN Ia-CSM SN 2020eyj at $\sim$1.5-2 yrs, we have carried out the largest late-time ($>$1 yr) radio survey of SNe Ia-CSM to date. The goal was to detect synchrotron emission from shock interaction with a high-density CSM, as suggested by optical observations, that would have eluded earlier observations due to heavy synchrotron and free-free absorption. 

We observed 29 historical SNe Ia-CSM with ages of 1--25 yrs with the Very Large Array for about 10--20 mins each, deep enough to detect emission at levels similar to SN 2020eyj. Images were calibrated and reconstructed with standard pipelines, and the flux densities (or upper limits) at the SN locations were analyzed with standard synchrotron emission models from SN shocks interacting with  $r^{-2}$-like CSM wind. We used a new model of the shock dynamics from \cite{TC17} that includes a Sedov-Taylor evolution of the shock if the swept-up CSM mass becomes comparable to or exceeds the ejecta mass. The new light curve model predicts increasingly fainter emission at higher mass-loss rates as a result of the enhanced deceleration of the forward shock.
 
Most of our SNe (28 out of 29) are radio non-detections down to 3$\sigma$ limits of $\lesssim$35 $\mu$Jy, implying that emergent radio emission (similar to SN 2020eyj) at these brightness levels on $1.5-25$ yr timescales is unlikely in SNe Ia-CSM. The only radio-detected object in the sample, SN 2022esa, is most likely a peculiar SN Ic based on inspection of spectra taken at later epochs.  Nevertheless, we have compiled the \emph{largest} sample of SNe Ia-CSM with radio upper limits in this paper, which allows us for the first time to independently verify the mass-loss rates of SNe Ia-CSM quoted from optical observations. We find that the radio upper limits of our sample rule out mass-loss rates between $\sim$10$^{-4}$ and 10$^{-2}$ \masslossunit (though the range differs for each SN and assumed model parameter). This is inconsistent with the mass-loss rates predicted by models of the H$\alpha$ line, yet consistent with the high mass-loss rates ($>10^{-1}$ \Msunyr) derived from bolometric light curve models of SNe Ia-CSM. We believe the best {bf way to reconcile radio and optical constraints is} either that the CSM has more complex, discontinuous geometries than a wind (e.g. shells that only produce transient H$\alpha$/radio emission), or that the CSM can be a wind, but the efficiency parameters of accelerating electrons ($\epsilon_e$) and amplifying magnetic fields ($\epsilon_B$) are much smaller for SNe Ia-CSM  ($<10^{-4}-10^{-3}$) than generally assumed for SNe Ia.

We hope the dataset presented here will be a valuable resource to the community for additional investigations of SNe Ia-CSM. %In follow-up papers, we will investigate the allowed properties of CSM shells given the upper limits from our observations \citep[following][]{Harris2021}. 
We particularly encourage coordinated optical and radio observations to better constrain CSM with discontinuous, shell-like geometries. The upcoming Rubin observatory, as well as high-sensitivity all-sky radio surveys with DSA-2000 \citep{Hallinan2019}, SKA \citep{Dewdney} and ngVLA \citep{Murphy2018} will not only enable such coordinated observations, but also reach greater sensitivities than our current observations.

\begin{acknowledgments}
\section*{Acknowledgments}
O.G., G.S., C.E.H., and L.C.\ are grateful for support from NSF grants AST-2107070 and AST-2205628. J.M. and M.P.T.\ acknowledge financial support through the Severo Ochoa grant CEX2021-001131-S and the Spanish National grant PID2023-147883NB-C21, funded by MCIU/AEI/ 10.13039/501100011033, as well as support through ERDF/EU. The National Radio Astronomy Observatory is a facility of the National Science Foundation operated under cooperative agreement by Associated Universities, Inc.
%This research is based on observations made with the NASA/ESA Hubble Space Telescope obtained from the Space Telescope Science Institute, which is operated by the Association of Universities for Research in Astronomy, Inc., under NASA contract NAS 5–26555. Funding acknowledgements for JWST, HST, Spitzer, Mayall? (Massey maps). These observations are associated with program(s) XXXXXXXXXX. This work is based [in part] on observations made with the NASA/ESA/CSA James Webb Space Telescope. The data were obtained from the Mikulski Archive for Space Telescopes at the Space Telescope Science Institute, which is operated by the Association of Universities for Research in Astronomy, Inc., under NASA contract NAS 5-03127 for JWST. These observations are associated with program XXXXXX. This work is based [in part] on observations made with the Spitzer Space Telescope, which was operated by the Jet Propulsion Laboratory, California Institute of Technology under a contract with NASA. 

\end{acknowledgments}


\vspace{5mm}
\facilities{VLA}


\software{\texttt{CASA} \citep{CASA}, \texttt{astropy} \citep{Astropy}, \texttt{matplotlib} \citep{matplotlib}, \texttt{scipy} \citep{scipy}, \texttt{numpy} \citep{numpy}, \texttt{photutils} \citep{photutils}.}


\appendix
\section{Comparison of Tang \& Chevalier (2017) with previous models} \label{app:A}

\begin{figure*}
    \centering
    \includegraphics[width=1\linewidth]{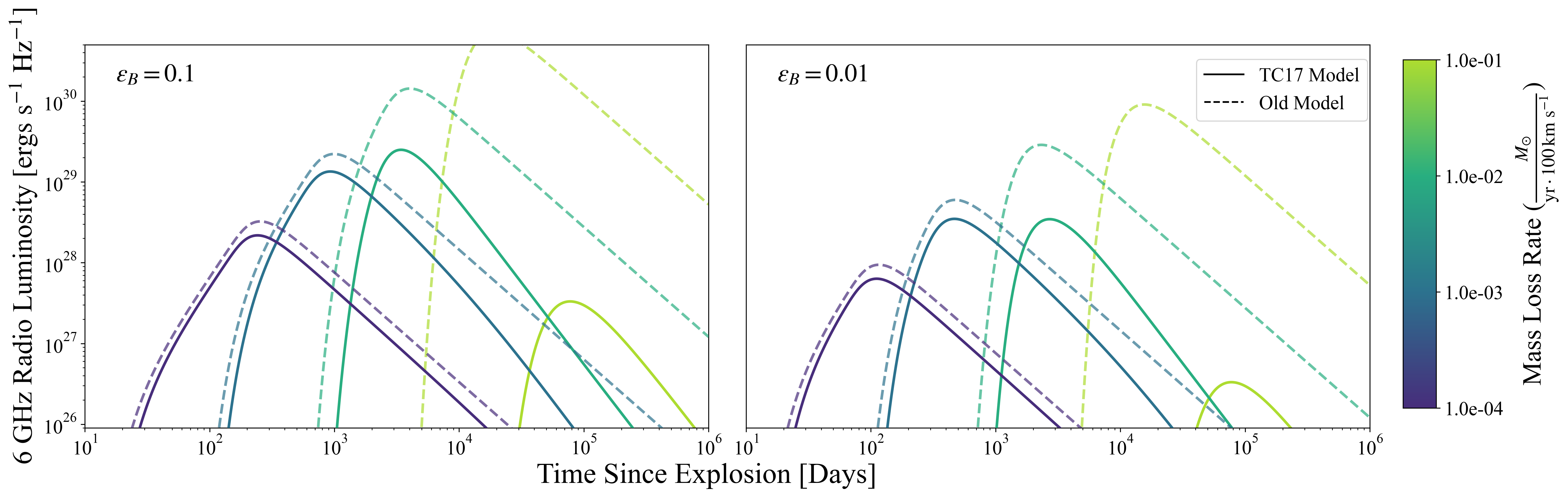}
    \caption{Comparison of light curves produced by different models. The solid lines make use of the dynamics from TC17, while the dashed lines represent the models of \cite{chomiuk16}, which do not include the transition to the ST phase. The left panel shows models for $\epsilon_B = 0.1$, and the right panel for $\epsilon_B = 0.01$. Different color lines represent different values of $\dot{M}/v_{wind}$, according to the color bar at right.}
    \label{fig:lccs}
\end{figure*}
\begin{figure*}
    \centering
    \includegraphics[width=1\linewidth]{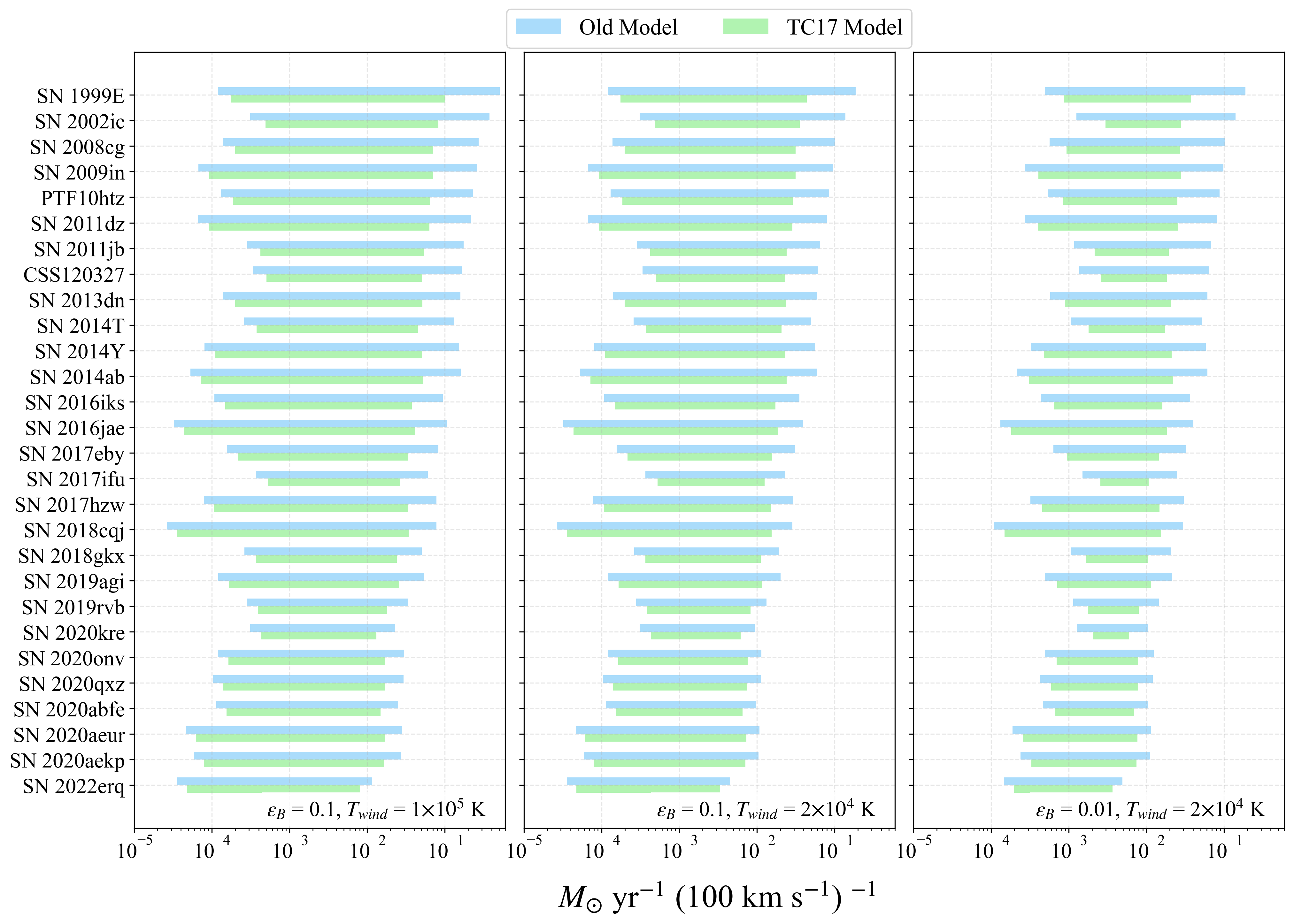}
    \caption{Comparing the TC17 model (green) with the older, ejecta-dominated light curve models (blue). Each panel represents the parameter value combinations from Table \ref{tab:params}. See Appendix \ref{app:A} for details. This figure does not include data for SN 2022esa.} 
    \label{fig:barh compare}
\end{figure*}

Here we discuss the difference between the light curves based on the shock dynamics of TC17 model that includes both ejecta-dominated and Sedov-Taylor evolution, and those assumed in previous radio studies of SNe Ia \citep[e.g.][]{chomiuk16} that only included the ejecta-dominated solution, $R \boldsymbol{\propto} t^{(n-3)/(n-s)}$. For our assumed $n=10$ ejecta and wind-like CSM ($s=2$), the ejecta-dominated solution implies $R_s \boldsymbol{\propto} t^{7/8}$ and $v_s \boldsymbol{\propto} t^{-1/8}$. While the ejecta-dominated solution is applicable to the general SN Ia population where CSM densities are expected to be low \citep{chomiuk16}, it can lead to very different light curves at high CSM densities, compared to those including the Sedov Taylor phase.

This effect is illustrated in Figure \ref{fig:lccs}, where we plot the ejecta-dominated+Sedov based light curve (the ``TC17 model'') in solid, and the ejecta-dominated-only light curves from previous models (``Old Model'') in dashed.  For lower mass loss rates ($<10^{-4}$ \Msunyr), the light curves for the two models are almost identical since the shock is ejecta-dominated in both models; the Sedov phase has not set in during the timescale of our observations. At higher mass-loss rates, the difference between the two models become more dramatic, with the TC17 model showing decreasing luminosities with increasing mass-loss rate, while the old models show increasing luminosities. The difference as mentioned in Section \ref{subsec:basic_properties} is because at high densities, the onset of the Sedov phase is occurring before the light curve becomes optically thin to free-free absorption. The Sedov solution has a higher rate of deceleration than the ejecta-dominated phase, so the decrease in the magnetic field and electron energy densities (that power the synchrotron emission) is occurring faster. By the time the light curve emerges from free-free absorption therefore, the luminosity has already significantly decreased compared to the ejecta-dominated case, giving an overall fainter light curve in the TC17 models.

In Figure \ref{fig:barh compare}, we see the difference between the two models in the range of mass-loss rates ruled out by observations. The difference in the lower end of the ruled out range between the two models is minor ($\sim$0.1 dex, the largest being 0.4 dex in SN 2002ic for $\epsilon_B=0.01$, $T_{wind}=2\times10^4$ K), caused mainly by the slightly steeper slope of the light curve in the ST phase for mass-loss rates approaching 10$^{-3}$ \masslossunit.  We see a more noticeable difference between the two models for the upper end of the ruled-out range  (with TC17 on average being 0.4 dex lower, the largest being 0.7 dex in SN 1999E). This difference gets larger for older SNe because the light curves of the ``old" model get brighter with mass-loss rate, and so a larger extent (in mass-loss rate) of the light curve can be excluded by the upper limit (Figure \ref{fig:2011dz}). Additionally, the time when $\tau_{\mathrm{ff}} \gtrsim 1$ is prolonged for the TC17 model because of the slower evolution of the shock, further restricting the upper limit that can be constrained. 

\clearpage
\bibliography{ogg}{}
\bibliographystyle{aasjournal}

\end{document}